\documentclass{article}

\usepackage[english]{babel}

\usepackage[a4paper,top=2cm,bottom=2cm,left=2cm,right=2cm,marginparwidth=2cm]{geometry}
\usepackage{comment}
\usepackage{amsmath}
\usepackage{graphicx}
\usepackage[colorlinks=true, allcolors=blue]{hyperref}
\usepackage{authblk}
\usepackage{hyperref}
\usepackage[hyphenbreaks]{breakurl}
\usepackage[figuresright]{rotating}
\usepackage{float}
\usepackage{multirow}
\usepackage{ulem}
\usepackage{todonotes}
\usepackage[title]{appendix}
\usepackage{longtable}
\usepackage[utf8]{inputenc}
\usepackage{graphicx,color,caption}
\usepackage{epstopdf}
\usepackage{geometry}

\title{\textbf{Deep learning forward and reverse primer design\\ to detect SARS-CoV-2 emerging variants}}
\author[1]{Hanyu Wang}
\author[2]{Emmanuel K. Tsinda\footnote{Currently affiliated with Center for Biomedical Innovation, MIT, Cambridge, MA 02139-4307, United States}}
\author[1]{Anthony J. Dunn}
\author[3]{Francis Chikweto}
\author[4]{Nusreen Ahmed}
\author[4]{Emanuela Pelosi}
\author[1]{Alain B. Zemkoho}
\affil[1]{School of Mathematical Sciences, University of Southampton, United Kingdom}
\affil[2]{Department of Virology, Tohoku University Graduate School of Medicine, Japan}
\affil[3]{Department of Biomedical Engineering, Tohoku University,
Japan}
\affil[4]{Specialist Virology Centre, University Hospital Southampton, United Kingdom}

\begin{document}
\maketitle


\begin{abstract}
Surges that have been observed at different periods in the number of COVID-19 cases are associated with the emergence of multiple SARS-CoV-2 (Severe Acute Respiratory Virus) variants. The design of methods to support laboratory detection are crucial in the monitoring of these variants. Hence, in this paper, we develop a semi-automated method to design both forward and reverse primer sets to detect  SARS-CoV-2 variants. To proceed, we train deep Convolution Neural Networks (CNNs) to classify labelled SARS-CoV-2 variants and identify partial genomic features needed for the forward and reverse Polymerase Chain Reaction (PCR) primer design. Our proposed approach supplements existing ones while promoting the emerging concept of neural network assisted primer design for PCR.
Our CNN model was trained using a database of SARS-CoV-2 full-length genomes from GISAID and tested on a separate dataset from NCBI,  with 98\% accuracy for the classification of variants. This result is based on the development of three different methods of feature extraction, and the selected primer sequences for each SARS-CoV-2 variant detection (except Omicron) were present in more than 95 \% of sequences in an independent set of 5000 same variant sequences, and below 5 \% in other independent datasets with 5000 sequences of each variant. In total, we obtain 22 forward and reverse primer pairs with flexible length sizes (18-25 base pairs) with an expected amplicon length ranging between 42 and 3322 nucleotides. Besides the feature appearance, in-silico primer checks confirmed that the identified primer pairs are suitable for accurate SARS-CoV-2 variant detection by means of PCR tests.

\end{abstract}


\section{Introduction}
The COVID-19 disease resulting from the Severe Acute Respiratory Virus (SARS-CoV-2) has caused a global pandemic since December 2019, affecting millions of humans \cite{Simonsen2021-vf,world2020novel}, and disrupting the medical and socio-economical systems \cite{doi:10.1073/pnas.2006991117}. The SARS-CoV-2 is transmissible by air-droplets or close contact with infected individuals \cite{nature2021coronavirus} and molecular virus detection is commonly done using nucleic acid amplification methods. SARS-CoV-2, like other RNA viruses, can mutate rapidly to generate variants that are associated with increased virus transmission, vaccine breakthrough, and higher COVID-19 detection rates \cite{doi:10.1021/acsinfecdis.1c00557, Christensen2021-da}. At present, high throughput sequencing or Next Generation Sequencing (NGS) is the reference laboratory method for SARS-CoV-2 variants identification \cite{10665-338480}. NGS requires a device and expert training to generate and analyze the data. Another sequencing method called Sanger sequencing approach is less expensive than NGS method. However, it is time consuming for genome sequences with 30 kilobase pairs like the SARS-CoV2 genome \cite{NGS_vs_Sanger}. PCR molecular detection assays are more affordable and customizable than sequencing workflows assays for the detection of known genes. Thus, PCR-based assays have been widely adopted for microbial detection in both research and clinical settings \cite{NGS_vs_qPCR, Johnson2021-dj}. The conventional PCR assay requires a target gene, a primer pair (forward and reverse), in addition to the polymerase enzyme, oligonucleotides and other salts in the reaction mix \cite{lorenz2012polymerase}. PCR primers must be designed appropriately to ensure the specific binding of primers to the known target sequence.

\subsection{Limitations of existing primer design methods}
As stated in  \cite{BUSTIN201719},  ``primers are arguably the single most critical components of any PCR assay.'' A reliable PCR test requires good primers, and the traditional approach of primer design contains forward primer design and reverse primer design, which starts by gathering similar sequences that harbor the intended target region. Sequence alignment allows for better visual screening and selection of the conserved 18-25 base pair regions that would cover the target region. Once a candidate primer is found, thermodynamic properties such as guanine-cytosine content (GC content), melting temperature, secondary structures should be assessed to ensure primer efficiency and minimize primer dimers. This stretch of oligonucleotides that is complementary to the anti-sense strand would be considered the forward primer. Besides the forward primer, a reverse primer must be designed downstream from the forward primer, following similar primer design rules, except that the reverse primer would bind to the sense strand of the target DNA. More details on this standard  process can be found in \cite{bustin2020parameters}.

However, the practical process is not as simple as described in theory by selecting sequences from a target gene, instead, it heavily relies on the expert knowledge of the operator in terms of sequence alignment, primer specificity and dimers checks. For example, to design primers for detecting the SARS-CoV-2 Alpha variant using the traditional approach, the operator needs to identify a genomic sequence of 18-25 nucleotides from a 30000-nucleotide sequence, which is unique to the Alpha variant and matching pre-defined thermodynamic criteria \cite{10.1093/bioinformatics/17.3.214}. This requires a great deal of time to carefully screen and compare the sequence with other SARS-CoV-2 variants and even other coronavirus species to ensure this stretch of oligonucleotides is unique to the Alpha variant only. With such complicated processes and significant time requirement and high level of concentration, mistakes are very likely, and can easily lead errors or other human impacts. Considering such potential instabilities, we propose a deep learning-based semi-automated primer design method, which can address many of the aforementioned  issues, leading to a simple and convenient primer design for the detection of SARS-CoV-2 emerging variants.

\subsection{Proposed method and outline of results}
In this paper, we propose a deep learning method to reliably design both forward and reverse primers. For forward primer design, we start by collecting gene sequences via a pre-processing step.  Subsequently, a deep Convolutional Neural Network (CNN) framework is propose to accurately separate the target SARS-CoV-2 variant sequences from other variant sequences. Then the filters of the first convolutional layer of this model are analyzed to extract flexible length sequences (18-25 base-pair) which checks the suitability of primers.

For reverse primers, we also use a CNN technique  for the design, but while first requiring a separate data pre-processing approach, which uses the forward primers to locate positions in the target variant gene sequences and only takes the downstream sequences as a new dataset. 
Moreover, another synthetic dataset of sequence with the same distribution of nucleotide of the downstream dataset is also generated as an input noise of the CNN. Then using a feature extraction method, the reverse primer resulting from this deep-learning aided design combined with the forward primer, as well as Primer-Blast \cite{ye2012primer} can then be applied to check primer specificity to minimize off-target amplifications. In-vitro testing is also used to confirm the primer efficiency for the SARS-CoV-2 variant detection. Our method considers SARS-CoV-2 variants as exemplars, but it is applicable for the primer designs to amplify target genes of any organism and identify features for sequence sub-classification.

Our method is probably the first in the literature that enables a deep learning-based flexible length primer design,
and more specifically capable to design exclusive primers for sequence sub-classification, such as each variant of SARS-CoV-2. As a result, our deep learning-based semi-automated primer design process can precisely generate both forward and reverse primer for PCR assay to detect SARS-CoV-2 variants for the first time. To be precise, our approach can classify each SARS-CoV-2 variant with over 98\% accuracy when applied to the data. For each variant-targeting primer sequence, our method presents in over 95\% sequences of target variants, and less than 5\% in other variants (except Omicron, for 80\% and 20\% separately). In addition, our primer pairs obtained through our method targeting the Alpha, Delta, and Omicron variants can generate amplicons of length below 200 bp, which have wider range of applications such as design a conventional PCR, RT-PCR or qPCR assay specifically targeting these variants afterwards.

\subsection{Related work}
In order to increase primer design efficiency, many researchers developed various tools to support primer design. Primer3 is one of the most commonly used open-source software to aid primer design and its initial version was released in 2000  \cite{krawetz2000bioinformatics, untergasser2012primer3}. Primer3 uses existing thermodynamic models to compute the melting temperatures of oligonucleotides when hybridized to their target,  and then predicts the stability of secondary structures. Although Primer3 can help identify suitable primer pairs for the species-level targets of SARS-CoV-2, it is is less efficient in generating many invalid primers. Also, it is generally unable to identify primer pairs following classification of variants of the same virus specie. Moreover, Primer3 limits the input sequence length  at up to 10,000 bps.
Therefore, other methods have been recently developed to provide a more reliable primer design. Finite state machines (FSMs) has been used to classify the design of PCR primers as good or bad \cite{ashlock2004training}; this approach effectively assists researchers to select good primers generated from Primer3. However, using this method needs a large amount of primers available as training dataset, and for viruses such as SARS-CoV-2, which can mutate rapidly, this method is less tractable to keep up with the pace of required updates.

A genetic algorithm to design  primers has been proposed in \cite{wu2004primer}; the method is not affected by the length of the sequence and could be much more precise and efficient than Primer3. However, it relies heavily on the simulation of random mutations, which can be unstable for SARS-CoV-2 virus with multiple mutation areas and the ability of rapid mutation, and it still cannot detect variants.
 Another evolutionary algorithms to design specific primer for SARS-CoV-2 variants can address the design of both forward and reverse primers for variants \cite{rincon2021design}. But similar to the genetic algorithm mentioned above, this approach relies heavily on mutation areas on sequence by selecting the most significant mutation to help the algorithm to find primers in this area, which would be unstable when mutation changed and not generic for other viruses because the selected mutation is fixed. Besides, if the mutation is unknown this method cannot be processed.

A machine learning method is developed in \cite{lopez2021classification} to design forward primer target SARS-CoV-2 virus, which allows the generation of forward primers and not limited by sequence length. However, this method can only handle a 21 base-pair fixed forward primer design while reverse primers still need to be selected manually. Also, it can only generate fixed length of primers.

Compared with traditional primer design and above mentioned approach, our method offers many advantages, such as the reduced human effort to identify unique genome regions in a file containing thousands of sequences with a 30000-nucleotide length. At the same time, our method is more efficient than Primer3 for the primer design of sequence sub-classification without the limitation of sequence length input, see Table \ref{tab:14}. Moreover, our method introduced machine learning-based reverse primer design for the first time, and there different feature extraction schemes are developed in this study. It can generate flexible length size primers (18-25 base pairs) without requiring other conditions than gene sequences of the virus to design the specific primers for the virus variants, making it simpler and more convenient for researchers to operate, even if for the virus that can mutate rapidly. In addition, as our method achieved significant sequence classification accuracy, this deep learning aided primer design would be very useful to design of primers for assaying other RNA viruses.

\subsection{Outline of the paper}
We start the next section with data collection from GISAID and NCBI dataset used in this project and pre-processing for forward primer design. Subsequently, we introduce the architecture of deep-convolutional neural network used in this study, following with the feature extraction and 3 different evaluation methods from CNN model to generate candidate forward primers. Section \ref{section_Reverse_primer_design} is devoted to reverse primer design, which is first proposed deep learning approach of reverse primer design, and validated using in-silico PCR software. Section \ref{section_Numerical_results} presents numerical results of this study including both the forward and reverse primer design, and covers primer pairs that designed for each SARS-CoV-2 variant validated by BLAST and in-silico PCR. Finally section \ref{section_Final_discussion} gives a brief conclusion of this paper and potential applications of this study are discussed.


\section{Forward primer design} \label{section_Forward_primer_design}


The forward primer design is the first step for our deep-CNN model  to generate candidate forward primers. This section will introduce our approach, going from data collection and pre-processing to feature extraction and evaluation, with the satisfactory forward primers of each SARS-CoV-2 variant of eventual output.

\subsection{Data collection and pre-processing} \label{Data collection and Pre-processing}

\paragraph{Data collection.}
The data used in this project comes from GISAID (\href{https://www.gisaid.org}{Global Initiative on Sharing Avian Influenza Data}, \cite{shu_mccauley_2017}) and NCBI (\href{https://www.ncbi.nlm.nih.gov}{National Centre for Biotechnology Information}, \cite{NCBI_webpage}). The data is  collected to 1) train the CNN model used for variant classification, 2) input into the trained CNN model to generate candidate forward primers, 3) obtain downstream sequences after the candidate forward primers used for reverse primer design, and 4) calculate the appearance of forward and reverse primers to ensure they only appear in the target variant.
The total number of used SARS-CoV-2 gene sequences from GISAID for training and validation are 473,645. The data set is comprised of five SARS-CoV-2 variants (Alpha, Beta, Gamma, Delta, and Omicron),  labelled based on the World Health Organization (WHO) nomenclature, the Pango lineage, and the GISAID clade. The number of samples and labels assigned in this project are listed in Appendix table \ref{tab:1}.

\paragraph{Data pre-processing.}
Sequence length-based filtration is performed by computing the average length of each variant sequence and eliminating sequences shorter than 2/3 of the average length. This procedure ensures completeness of all the genomes in our data set where the longest virus genome (31,079 bps) is identified and recorded. Appendix table \ref{tab:3} shows the average sequence length of each variant. To ensure the consistency of genome sequences in the data set, the letter $N$ is incorporated into other genome sequences resulting uniform in size and treating as a $1 \times 1 \times 31,079$ matrix (or vector, more precisely).

Deep learning convolutional neural models need encoded data in numeric variable rather than categorical data like base pairs ``A'', ``T'', ``C'' or ``G'' \cite{zhang2021deep}. As a result, before each base in the gene sequence being fed into the model, it is common to use the following special function to encode a nucleotide:

\begin{equation} \label{eq:1}
Y := f(x) := \left \{
\begin{aligned}
    0.00 & \quad \mbox{ if } & x=N,\\
    0.25 & \quad \mbox{ if } & x=C,\\
    0.50 & \quad \mbox{ if } & x=T,\\
    0.75 & \quad \mbox{ if } & x=G,\\
    1.00 & \quad \mbox{ if } & x=A.\\
\end{aligned}
\right.
\end{equation}
To ensure the readability of the model output, an inverse of the above function converts the output numeric results back to letters. Similarly, for the SARS-CoV-2 variants, different label data are onne-hot encoded by converting the labels into a $1\times 5$ vector.
\begin{table}[H]
\centering
\begin{tabular}{c|ccc|cc}
 & Training set & Validation set & Test set & Generated primers & Calculated appearance\\ \hline
Source & GISAID & GISAID & NCBI & GISAID &  GISAID and NCBI \\ \hline
Alpha & 2,000 & 2,000 & 2,000 & 1,000 (or) & 5,000 \\
Beta & 2,000 & 2,000 & 2,000 & 1,000 (or) & 5,000 \\
Gamma & 2,000 & 2,000 & 2,000 & 1,000 (or) & 5,000 \\
Delta & 2,000 & 2,000 & 2,000 & 1,000 (or) & 5,000 \\
Omicron & 2,000 & 2,000 & 2,000 & 1,000 (or) & 5,000 \\
Other Taxa & 0 & 0 & 0 & 0 & 3,640 \\ \hline
Total Number & 10,000 & 10,000 & 10,000 & 1,000 & 28,640 \\ \hline
\end{tabular}
\caption{Data selection for the CNN model, generated and calculated appearance rate of forward primers.}
\label{tab:4}
\end{table}

Reshuffling the sequences order in the dataset with selecting 4,000 sequences from each variants, a total of 20,000 sequences in a 1:1:1:1:1 ratio are separated into the training set and validation sets, while each set is randomly split into 200 small data sets with a batch size of 50 to avoid memory overflow \cite{he2019control} and bias in the results \cite{schuster2019towards}. The test set is created using sequences from NCBI database to aid in the evaluation of the primers specificity. However, since NCBI does not provide classification labels for SARS-CoV-2 variants, each variant label is assigned using PANGOLIN (Phylogenetic Assignment of Named Global Outbreak Lineages) \cite{10.1093/ve/veab064}.

To generate forward primers, 1,000 randomly selected sequences of each variant are analyzed (Table \ref{tab:4}). The appearance rate of generated primers is evaluated by the proportion of primers matching with sequences within independent datasets. These independent data from GISAID and NCBI comprise genomic sequences of various human and non-human coronaviruses (check Table \ref{tab:2} for details).

\subsection{Deep-convolutional neural network architecture}
\begin{figure}[H]
  \centering
  \includegraphics[width=1.0 \textwidth]{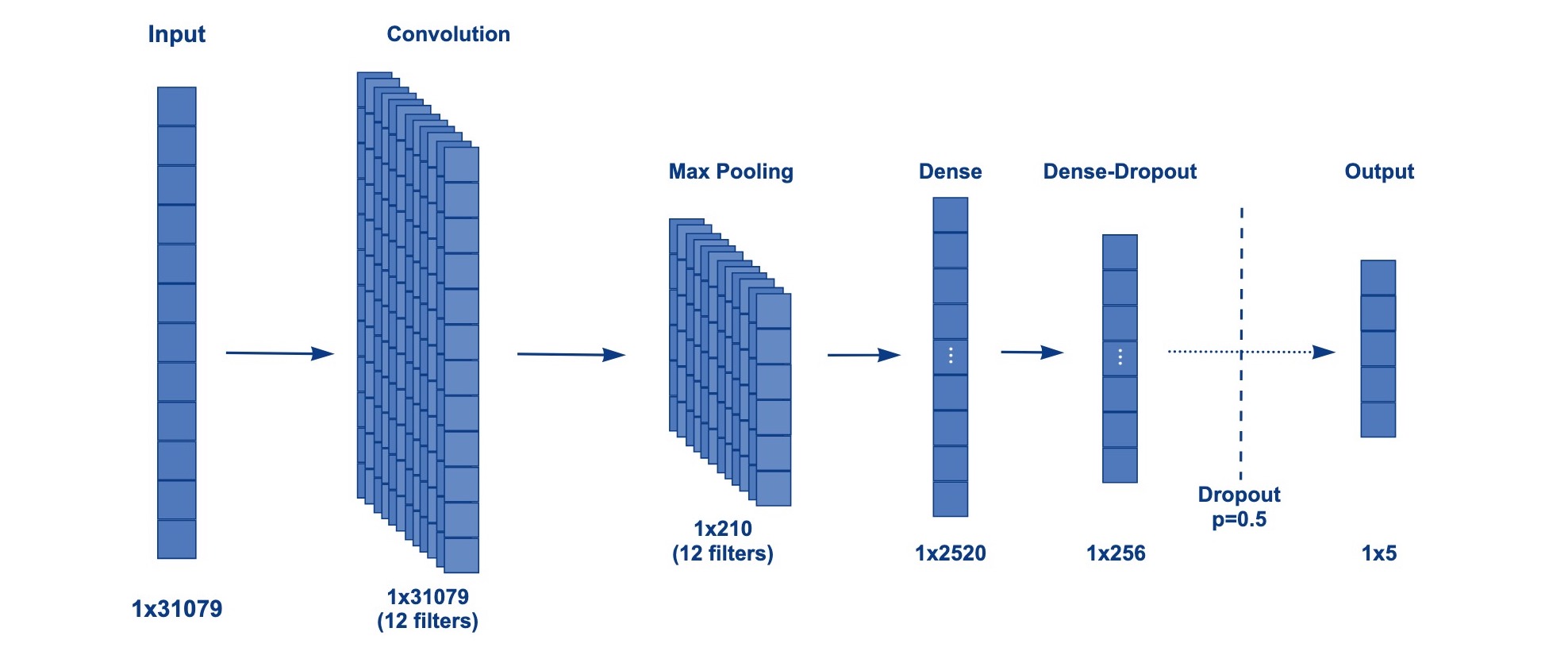}
  \caption{\label{fig:CNN_architecture_forward} A visual representation of the proposed architecture of the CNN model for forward primer design.}
\end{figure}
The deep-CNN architecture considered for the forward primer design is structured into three main layers. That is, the convolutional layer with a non-linear activation function, the MaxPooling layer, and the dense layer \cite{wu2017introduction} (see Figure  \ref{fig:CNN_architecture_forward}).
In the pre-processing stage, sequences of SARS-CoV-2 variants are standardized into a format of the same size (31079 in this project), and all sequences are considered as a $1 \times 1 \times 31079$ matrix, which will be an input to the CNN model. The convolutional layer of CNN consisted of 12 filters with $1 \times N$ kernel size, where N equals to the length of the primer finally generated. After the data being fed into the model, the convolution operated by the 12 filters, and element-wise non-linearity is applied to the generated feature maps from the convolution kernels using Rectified Linear Unit (ReLU) \cite{nair2010rectified} to accelerate the convergence of the CNN model. As a result of performing convolution operations on the input data, a volumetric feature map of size $1 \times 31079$ with depth 12 is generated.

To reduce the dimension of the feature maps, maximum pooling is applied. While selecting the ideal maximum pooling window size for the pooling layer, sizes of $1\times 75$, $1\times 150$, and $1\times 300$ are tested. The $1 \times 150$ window size showed the best performance. Eventually, a $1 \times 148$
pooling window size in pooling layer is implemented, and the resulting feature map size after pooling reduces to $12 \times 1 \times 210$.
It should be noted that the dense layer (fully connected layer) is responsible for data collation of the feature maps produced from convolutional layer \cite{BASHA2020112}. Additionally, the $12 \times 1 \times 210$ feature map is flattened to a $1 \times 2520$ matrix and subsequently reduced to a matrix of $1 \times 256$ for the  preparation of the final output. Popular CNN models such as YOLOv1 \cite{redmon2016you} usually adopt two fully connected layers. In this project, we use a second fully connected layer added with a 50\%  dropout probability that reduces the $1 \times 256$ data to $1 \times K$, where K = 5 corresponds to the number of classes.
Furthermore, for the best classification results, we consider to use the Adam optimization algorithm \cite{kingma2014adam} with the loss function based on the softmax cross entropy  \cite{softmax_cross_entropy} to train the model parameters as a hyperparameter selection. Compared with other SGD (Stochastic Gradient Descent) optimizers, the Adam approach can adaptively learn weights which shows better results.

\subsection{Extracting and evaluating}
This built CNN model is trained to identify five types of the SARS-CoV-2 variants with a classification accuracy over 98 \% (Figure \ref{fig:TSNE_CNN_Classification}). After the classification step, we develop three novel feature extraction methods that require separation of the $12 \times 1 \times 31079$ feature maps into 12 layers by channels and saving each layer as a file respectively labelled from Filter\_0 to Filer\_11.

\begin{figure}[]
\centering
\includegraphics[width=0.7 \textwidth]{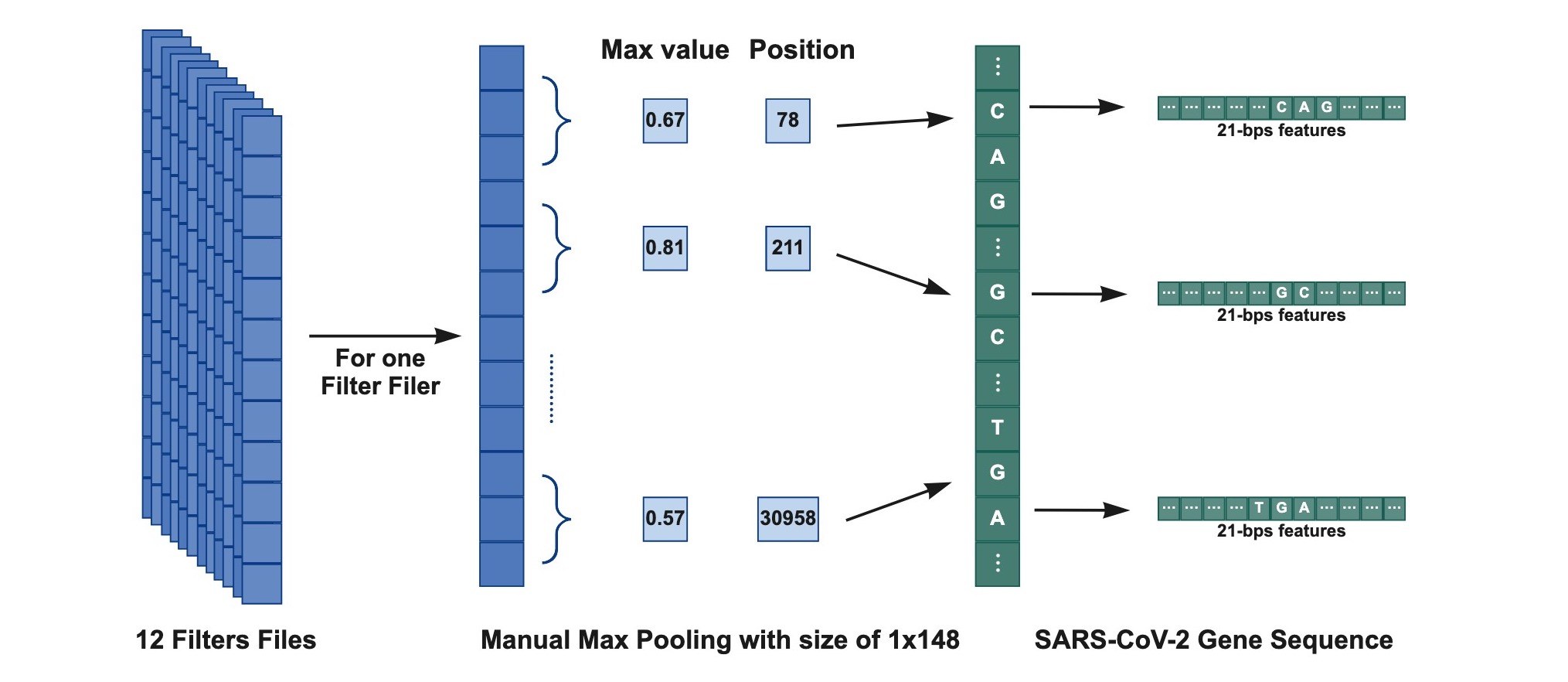}
\caption{\label{fig:Filter_maxPool}Feature extraction from the filter files by simulating  maximum pooling operation.}
\end{figure}

\paragraph{Pooling method.}
This approach is similar to the maximum-pooling operation in MaxPooling layer. The reason why we do not directly use the output of the MaxPooling Layer is because we do not exactly want the maximum value but need the positions of the maximum data in each pooling region. With this pooling method, each filter file is manipulated a manual maximum-pooling as shown in Fig. \ref{fig:Filter_maxPool}, and the positions of max value are recorded and saved as a PosPool file, and this process must be done individually for each filter file.

\paragraph{Top method.}
Unlike the pooling method, the top one does not need a pooling window size but to set a fixed variable called top variable as it is designed to find the maximum data in whole region of the layer rather than in each pooling region. While processing this approach, only a fixed number of max value positions that equals to top variable would be recorded, which means using different top variable the number of maximum data selected from the whole feature map would differ. With testing the top variable from 75, 125, 175, and 250 in Alpha and Delta variants, we obtain the results shown in Appendix table \ref{tab:10} with the number of generated candidate primer pairs. Finally, 175 is found as the most efficient top value after comparing different top variables.

\paragraph{Mix method.}
The mix method combines the pooling method and the top method. This method needs both a pooling window size and a top variable, which would process pooling operation first, then finds top maximum values in each pooling region and records the position of each top value. Refer to the test results of above two methods, we found that the most effective parameter setting is to use the maximum pooling window size of $1 \times 500$ and to both record and save the top 10 positions of the maximum data in each pooling region.

\hspace*{\fill}

After using above feature extraction method to create the PosPool files, each PosPool file is then combined with the input sequence data for the purpose of extracting the features associated with each variant. By using the positions of max value recorded in the PosPool file, we returns back to the input gene sequence for the nucleotides corresponding to the positions and expanded with this nucleotide as the centre to generate candidate primers with a flexible length size of 18-25 base-pair. For example, when we want to obtain a primer of length 25 bps, we take 12 nucleotides in forward and 12 nucleotides in postward of the selected central nucleotide combined together as a new 25 base-pair primer sequnence. Since this primer sequence is generated at the position corresponding to one of the largest numerical elements in the convolutional layer which aimed for variants classification. Hence, this central nucleotide contains feature information to distinguish specific variants and enables potential primer design. Nevertheless, these identified primer sequence are not considered as forward primers until they have met predefined primer design criteria \cite{dieffenbach1993general, primerdesign} that include:

\begin{itemize}
    \item []
    \begin{enumerate}
        \item Length of 18-25 nucleotides;
        \item 40-60 \% GC content;
        \item Start and end with 1-2 G/C pairs;
        \item Melting temperature (Tm) of 50-60$^{\circ}$C (using on Primer3; also see \cite{BasqueTm});
        \item Primer pairs should have a Tm difference within 5$^{\circ}$C;
        \item Primer pairs should not have self-complementary regions of more than 5 pairs (based on IDT oligoanalyzer \cite{IDTHomoDimer}).
    \end{enumerate}
\end{itemize}

We compute the appearance rate to ensure that candidate forward primers appear only once in target SARS-CoV-2 variant genome; since failure to appear or multi-appears in the target sequence may cause mis-priming, leading to PCR failure \cite{SCHRICK20161}. Moreover, the desired primer pair should  specifically target each variant of the SARS-CoV2 which is expected the primer pair will only appear in the genome sequence of the corresponding variant with a rate greater than 95 \% (except for Omicron, which uses 80 \%), but in other variants or in non-SARS-CoV-2 organisms, it should appear at a rate less than 5 \% (ideally 0 \%). The appearance rate is computed based on the data shown in the Table \ref{tab:4} in Section \ref{Data collection and Pre-processing}) and Appendix table \ref{tab:2}. The forward primers' overall frequency of appearance in target variant is higher than 95 \%, and lower than 5 \% in others, see Table \ref{tab:5} and Table \ref{tab:6}, \ref{tab:7} in Appendix, and we refer to them as satisfactory forward primers. These satisfactory forward primers will then be used for data pre-processing in the reverse primer design. Fig \ref{Flowchart_Forward} in Appendix shows the process to design forward primer.


\section{Reverse primer design} \label{section_Reverse_primer_design}

In the traditional method of primer design, the reverse primer start location is chosen downstream the forward primer. For example, if amplicon length of 200 base-pairs is desired, the reverse primer would be located 200 base-pairs downstream of the forward primer. Based on this concept, we have created a new reverse primer design method using deep learning to generate reverse primers.  Unlike the previously employed pre-processing method for forward primer design, only the satisfactory forward primers and the corresponding variant gene sequences would be used.

\subsection{Data pre-processing for generating the downstream dataset}
\begin{figure}[htp]
    \begin{minipage}{0.58\textwidth}
        \centering
        \includegraphics[scale=0.151]{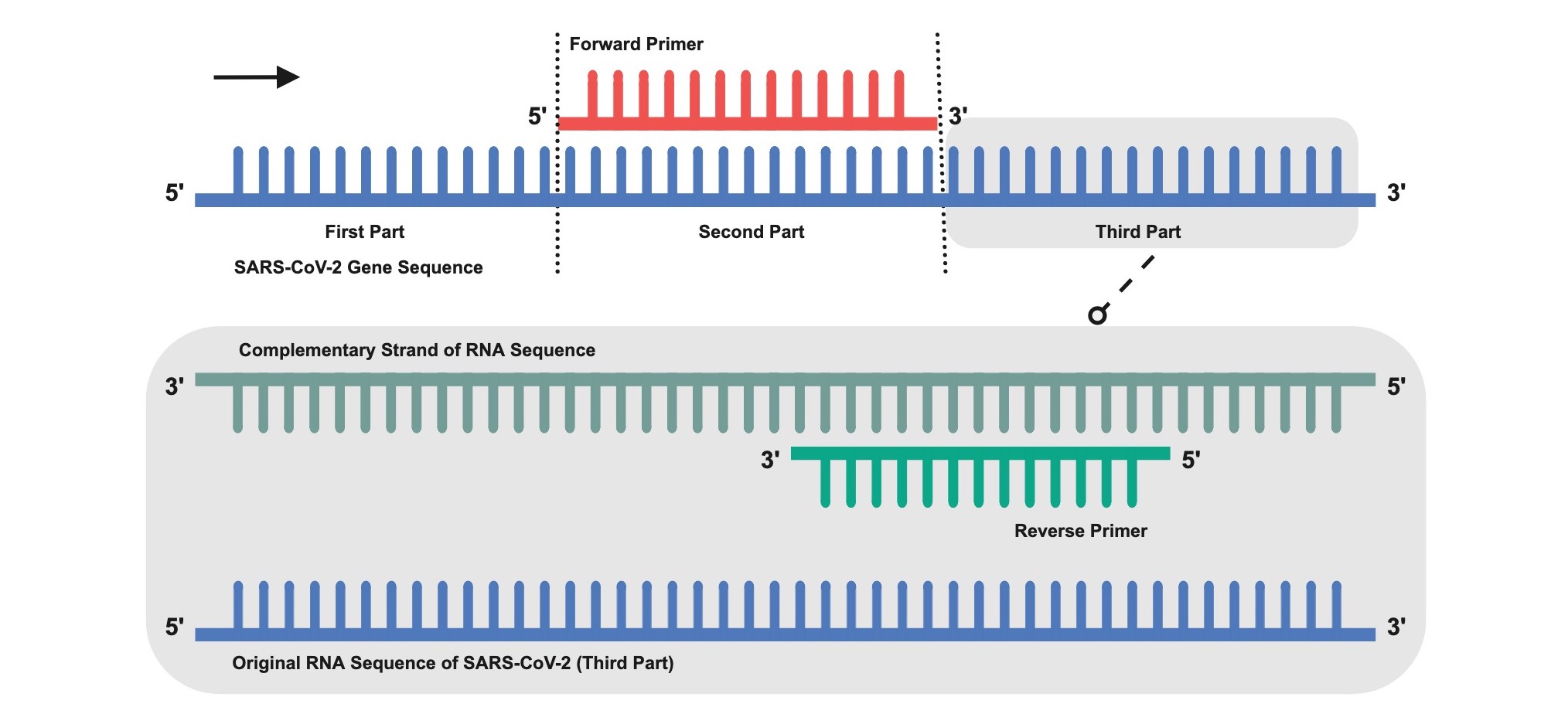}
        \caption{Separation of the gene sequence into 3 parts\\
        for reverse primer design. Note that  the positions and\\ length of the third-part sequences will differ for each\\ satisfactory forward primers. This is because
        the third-part \\
        would be unique and depending on the forward primer.}
        \label{fig:Third_Part}
    \end{minipage}
    \begin{minipage}{0.4\textwidth}
        \centering
        \includegraphics[scale=0.251]{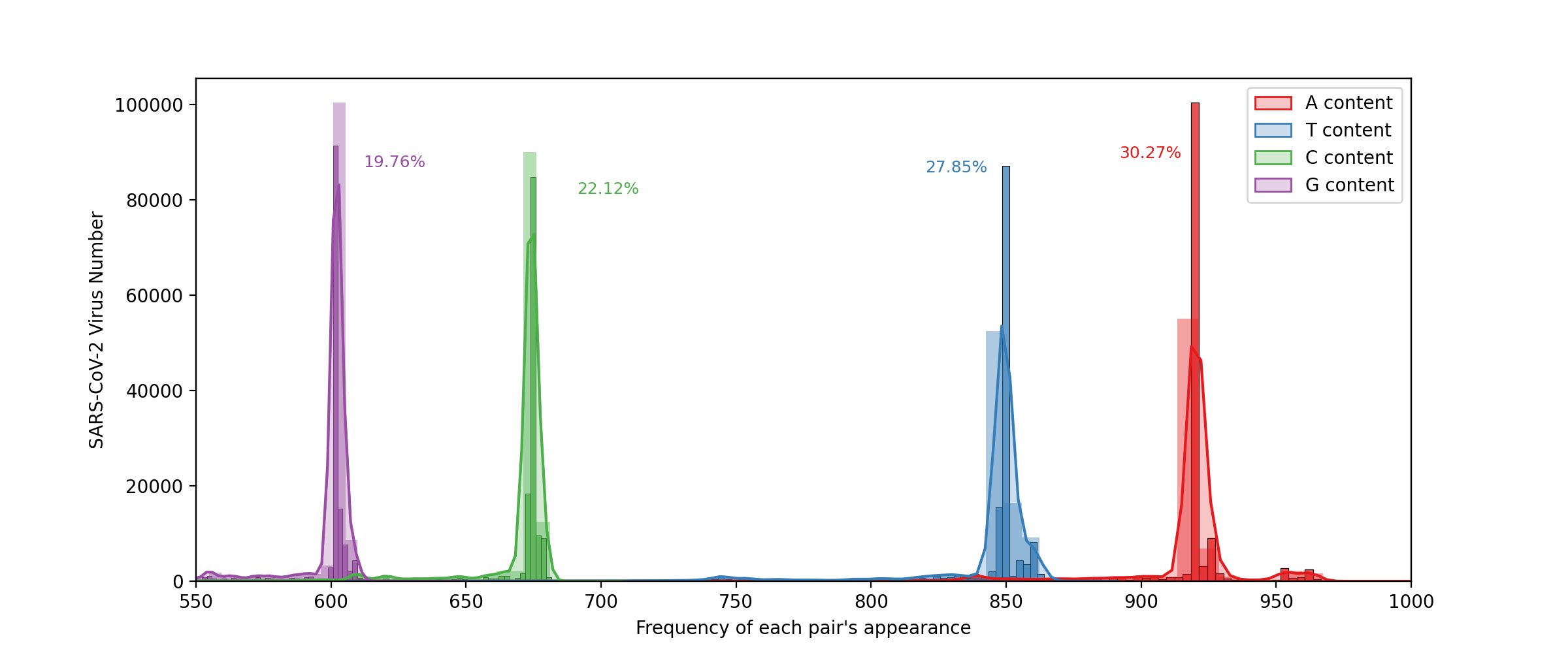}\caption{The third-part analysis of gene\\ sequences for Delta variant.}
        \label{fig:ATCG_content}
    \end{minipage}
\end{figure}

To prepare the data for reverse primer design only the target variant sequences and satisfactory forward primers are needed. The first step is to locate the position of satisfactory forward primer, and analyze the potential area of reverse primer starting from the satisfactory forward primer sequence till the end of the sequence. Considering the gene sequence is ordered from 5' to 3’ direction, the position of the reverse primer in the gene sequence should be after the forward primer and closer to the 3’ end. Next, trimming the collected full length variant-genome sequences containing the forward primer into 3 parts (see Fig \ref{fig:Third_Part}): (1) the sequence starting from the 5' end and ending before the forward primer first nucleotide (upstream); (2) the forward primer; and (3) the remaining sequences starting from the forward primer to the 3' end (downstream). Since the reverse primer would only appear in the third part where this part should be be preserved as the downstream dataset for the reverse primer generation.

\noindent
\\
\noindent
\subsection{Creation of the synthetic downstream dataset}
CNN model implemented for forward primer design is utilized in the same manner to design reverse primers, but currently only the third-part data that downstream to the satisfactory forward primer's binding location (see Fig \ref{fig:Third_Part}) for one variant is regarded as input to the CNN model. This leads to sequences with only one label for classification being not informative, and the accuracy of the model has no referential value, resulting in poor feature extraction as every solution for loss function would be the optimal solution. Therefore, it is necessary to add the other label for reference through create the synthetic downstream dataset by analyzing the current downstream dataset of each sequence. Calculating the ATCG content, then producing random sequences with the same distribution of each base, and inputting them into the model for training with the other label. Considering the Delta variant with satisfactory forward primer 'CTACCGCAATGGCTTGTCTTG', as an example, we initially select only the Delta variant genome sequences, and keep the downstream dataset of sequences obtained from the pre-processing step. Subsequently, analyze nucleotide distribution see Fig \ref{fig:ATCG_content} and create the synthetic downstream dataset data for reverse primer design.

\subsection{In-silico PCR}
In-silico PCR is a simulated or virtual representation \cite{lexa2001virtual} of the Polymerase Chain Reaction (PCR) useful to compute the theoretical PCR. It performed after the reverse primer design, to check whether the primer pair would specifically amplify the target gene. In this study, a similar manner as the forward primer design using CNN model is presented in reverse primer design with top method for feature extraction, then performing in-silico PCR using FastPCR \cite{kalendar2017fastpcr}, Unipro-Ugene \cite{ononechninov28ugene} and primer-BLAST \cite{ye2012primer} to validate the primer pairs initially considered satisfactory. Check Appendix fig \ref{Flowchart_All} for the overall workflow, and the specific process of reverse primer design, including in-silico PCR can be seen in Appendix fig \ref{Flowchart_Reverse}.


\section{Numerical results} \label{section_Numerical_results}
\subsection{Forward primer design}
\begin{figure}[htp]
    \begin{minipage}{0.5\textwidth}
        \centering
        \includegraphics[scale=0.47]{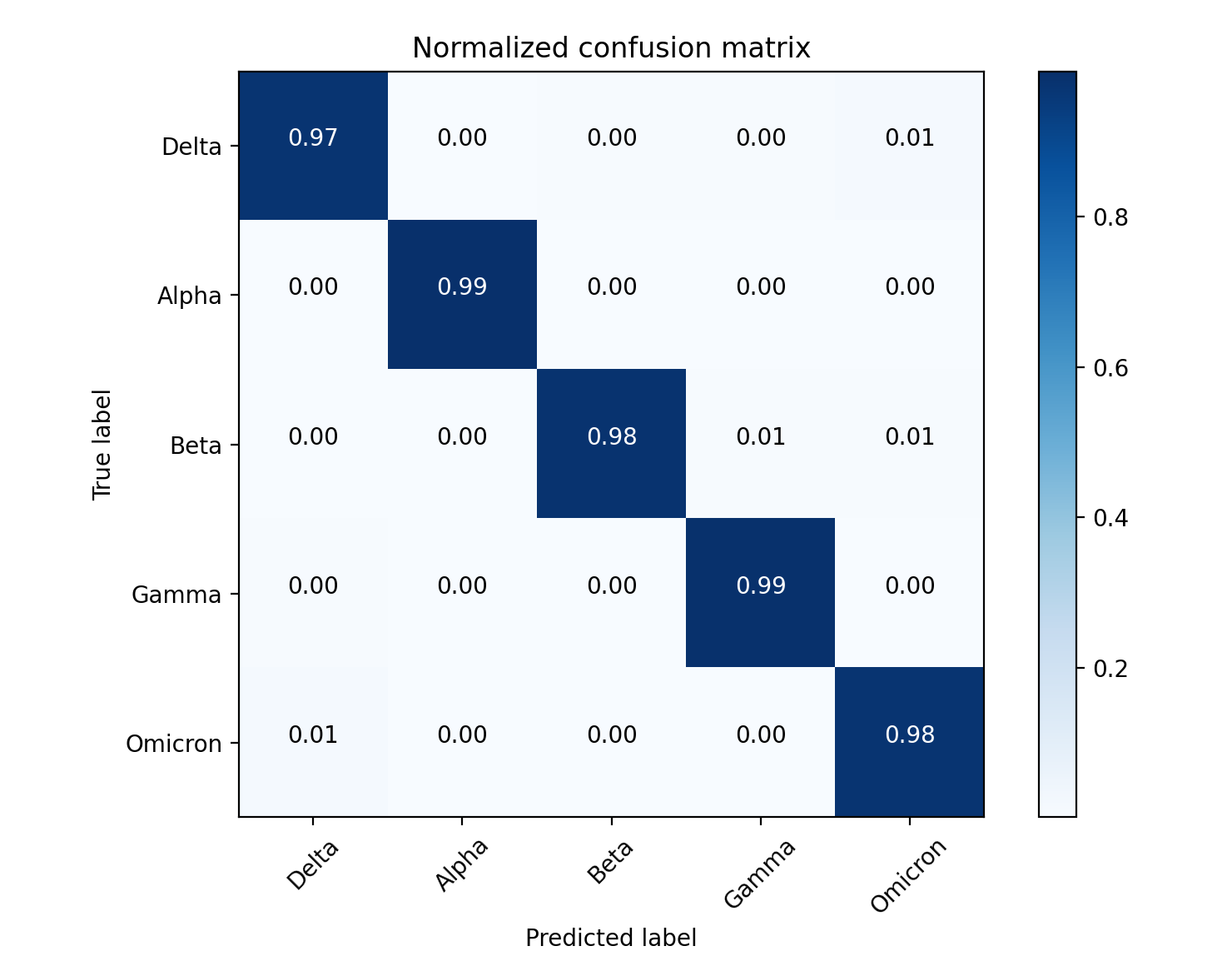}
        \caption{Confusion matrix from  cross-validation\\
        for the CNN model based on 2,000 sequences\\
        of each SARS-CoV-2 variant.}
        \label{fig:Confusion_Matrix}
    \end{minipage}
    \begin{minipage}{0.5\textwidth}
        \centering
        \includegraphics[scale=0.56]{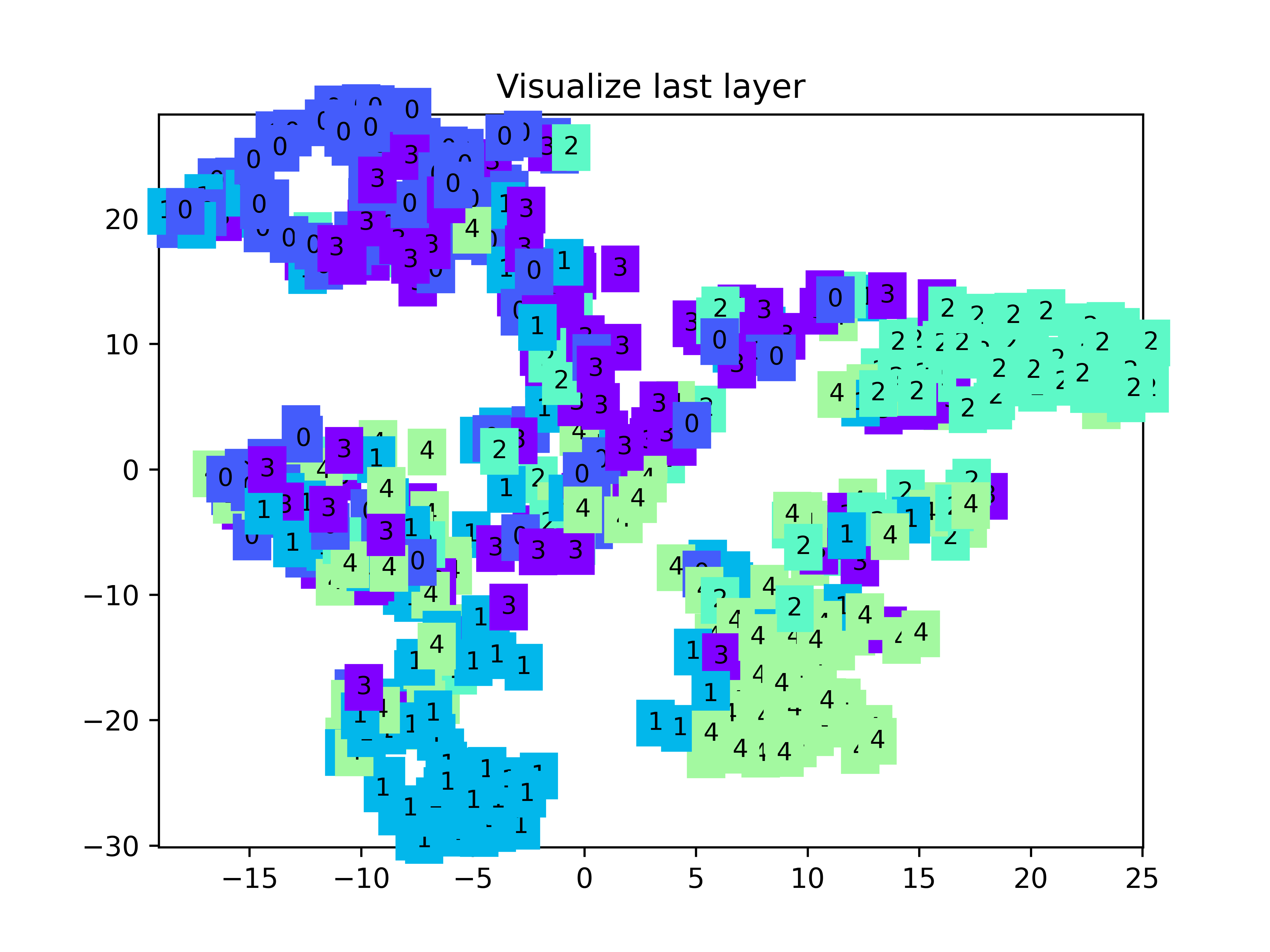}\caption{Classification results of five\\ SARS-CoV-2 variants after the model training.}
        \label{fig:TSNE_CNN_Classification}
    \end{minipage}
\end{figure}

The 610,000 full-length SARS-CoV-2 genome sequences retrieved from GISAID and NCBI database are split into a training set for the CNN model and a validation set of 10,000 variant sequences (comprised of 2000 sequences of each variant). Over 98 \% of the classification accuracy of the SARS-CoV-2 variants on both validation and test sets is obtained after the CNN model is trained for the forward primer design. Figure \ref{fig:Confusion_Matrix} shows the classification confusion matrix, while Figure  \ref{fig:TSNE_CNN_Classification} displays a summarized
visualization of the last layer.

The standardized genome sequences are passed through the convolutional layers and 12 filter files are generated. Different extraction methods are conducted in an experiment and similar results are obtained for both Alpha and Delta variants. The pooling method is chosen based on less time consumption and the number of forward primers generated comparing with other 2 methods. Screened by primer design criteria and frequency of appearance that resulting in the generation of 66 (Alpha), 23 (Beta), 59 (Gamma), 52 (Delta) and 69 (Omicron) satisfactory forward primers for each SARS-CoV-2 variant. The appearance rates using the Homo Sapiens (human) genome as reference can be seen in Table \ref{tab:5} in the appendices. While for non-Homo Sapiens hosts can be seen in Table \ref{tab:6}. The percentage of appearance of other taxa can be seen in Table \ref{tab:7}.
\noindent
\\
\noindent

\subsection{Reverse primer design} \label{Reverse Primer Design}
The CNN approach for the forward primer design is implemented with significant adjustments to design the reverse primer. The sequence data used for the forward and reverse primer design are different. The former includes sequences of all variants, while the latter one only uses sequences of one special variant that the forward primer targets. And we segmented these sequences associated with each forward primer into independent sets, then together with the synthetic dataset which input into CNN models after transforming to generate the reverse primer candidates. To complete this approach, we traine CNN models on separate dataset of each satisfactory forward primer, and test different feature extracting methods for reverse primer design (pooling, top and mix) with number of candidate reverse primer pairs obtained, as results shown below.

This experiment has been made in aim to compare the number of reverse primers generated in three extraction methods, see the result for Alpha variant on Appendix table \ref{tab:11} and for the result of Delta variant please check Appendix table \ref{tab:12}. From the result of Alpha variant, both top and mix method present a efficient feature extraction where pooling method works badly, while for Delta variant, top method has the most outstanding performance. Hence, finally decided to use the top method for feature extraction in reverse primer design. Then, we calculated the percentage of appearance for obtained reverse primer targeting each variant which can be seen on Appendix table \ref{tab:5.5}.

thermodynamic properties such as GC content and melting temperature (Tm) are checked, then self-dimer analysis is performed to exclude primers with a high number of self-complementary bases. Additional dimer analyses are performed to further exclude primer pairs with high number of complementary bases between the candidate forward and reverse primer. Besides, the difference in melting temperature (Tm) between primers pairs is kept within 5$^{\circ}$C. Based on the selection criteria mentioned above, 1,478 primer pairs in total were generated for all 5 types of variants, see Table \ref{tab:13}.

\begin{table}[H]
\centering
\small
\begin{tabular}{c|c|c|c|c|c}
 & Forward Primer  & Reverse Primer & Amplicon Size & Amplicon Size & Amplicon Size \\
 & Number & Number & \textless 200bps & \textless 500bps & \textless 1,000bps \\ \hline
Alpha & 66 & 400 & 6 & 14 & 66\\ \hline
Beta & 23 & 18 & 0 & 0 & 0 \\ \hline
Gamma & 59 & 272 & 0 & 49 & 66 \\ \hline
Delta & 52 & 457 & 33 & 106 & 154 \\ \hline
Omicron & 69 & 331 & 23 & 26 & 50 \\
\hline
\end{tabular}
\caption{Number of forward and reverse primers generated for each variant with amplicon interval.}
\label{tab:13}
\end{table}

\subsection{BLAST and in-silico PCR for primer validation}

\paragraph{BLAST}

Primer-Blast \cite{ye2012primer} analysis is performed following the computational design and testing for suitability as primer pairs. Although the appearance frequency of the primers in SARS-CoV-2 virus and other coronavirus species has been previously checked, we perform an additional Primer-BLAST analysis to confirm that the selected primer pair would specifically target SARS-CoV-2 virus, Appendix fig \ref{fig:BALST} and \ref{fig:BLAST_result} show the BLAST search parameters and results.

\paragraph{In-silico PCR}

The pre-selected genome sequences considered as potential primer pairs are validated by in-silico PCR testing using FastPCR \cite{kalendar2017fastpcr} and Unipro UGENE \cite{ononechninov28ugene}. The promising forward and reverse primers are entered into the in-silico PCR software in batches, and the in-silico PCR testing is performed. As a result, primer pairs obtained using our method are proven able to accurately detect SARS-CoV-2 variants as expected. When the primer pair is found functional, the in-silico PCR software specified the position of both primers within the target genome sequence, displaying the melting temperature (Tm) and the percentage of match with the target, as well as the amplicon size, see results in Appendix fig \ref{fig:FastPCR} and fig \ref{fig:Unipro_UGENE}. Eventually, we give a list of 22 validated primer pairs validated with best outcome by in-silico PCR is shown in Table \ref{tab:8}.

\section{Final discussion} \label{section_Final_discussion}

Our proposed deep learning assisted primer design approach provides a unique framework to design primers for SARS-CoV-2 variant detection, as it has been otherwise very challenging to find PCR primers sets for SARS-CoV-2 variants due to the genome heterogeneity across variants. Previously designed primers targeting SARS-CoV-2 variants are designed by manually screening hundreds of thousands of SARS-CoV-2 full length genomes looking for variant specific alterations such as deletions \cite{vogels2021multiplex} or variants specific mutations \cite{sibai2022development, wang2021multiplex}. This manual screening of genomes is extremely time consuming and requires human resource with advanced primer design skills. Also, such complex primer design may not be achieved satisfactorily by only using existing automated online tools such as Primer3plus \cite{untergasser2007primer3plus, untergasser2012primer3}, which requires input sequence length below 10,000 base pairs (Appendix table \ref{tab:14}). Thus, semi-automated deep learning aided primers design helps to decrease human efforts and human error when manually searching for SARS-CoV-2 variants specific features, see Table \ref{tab:14}, providing  a comparative experiment based on Delta Variant : hCoV-19/Indonesia/JK-GS-FKUINIHRD-0489/2022.

In-silico PCR validation shows that primer pairs that resulted from our semi-automated primer design would be effective for SARS-CoV-2 variant detection. Our deep learning models can scan through genome sequences and capture underlying features, making it possible to automate sequence classification for forward and reverse primer design. This is not only a new way to combine deep learning with traditional biology, but also, a practical use case of machine learning to enable a better understanding its potential future applications in genomics.

While our deep learning aided primer design algorithm is more efficient than the existing  tools (See Appendix \ref{Appendix_3}) for finding primer candidates for genome sequences, it does not incorporate degenerate oligonucleotides into the primer sequence. Introducing such a feature would mark another progress step in the whole area of automated primer design. The deep learning assisted primer design method still requires the primer candidate evaluation (Dimers, GC content, melting temperature, etc.), prior to selecting the best primer pair for in-vitro PCR testing. Additional research efforts towards this goal will enhance the potential of our CNN aided primer design method. Considering the significant  time requirement during model training, as well as the need for the algorithm to build a separate model for each forward primer, as well the training of the parameters for the reverse primer design, there is still space to improve our method.

\section*{Acknowledgement}

The work of Anthony J. Dunn is jointly funded by Decision Analysis Services Ltd and EPSRC through the studentship with Reference EP/R513325/1. The work of Alain B. Zemkoho is supported by the EPSRC grant EP/V049038/1 and The Alan Turing Institute under the EPSRC grants EP/N510129/1 and EP/W037211/1.

\section*{Conflict of interest statement}

The authors declare that there is no conflict of interest.

\section*{Data availability}
We gratefully acknowledge the authors responsible for obtaining the specimens and genetic sequence data generated and shared via the GISAID Initiative (\href{	https://doi.org/10.55876/gis8.220628xf}{https://doi.org/10.55876/gis8.220628xf}), and that we used for the research presented in this paper. All the codes used and generated in the course of the work presented in this article are available in the following GitHub page: \href{https://github.com/awc789/variant-primer-sars-cov-2}{https://github.com/awc789/variant-primer-sars-cov-2}

{\scriptsize{
\begin{center}
\begin{table}[H]
\centering
\begin{tabular}{c|c|c|c|c}
Primers (5' to 3') & GC content & Tm ($^{\circ}$C) &  Position & Amplicon Size\\ \hline
Alpha Variant \\ \hline
F - AGGAGCTATAAAATCAGCACC & 42.86\% & 49.60 & 27680-\textgreater27770 & \multirow{2}*{78 bps} \\
R - TCGATGCACTGAATGGGTGAT & 47.62\% & 53.89 & 27737\textless-27757 & ~  \\ \hline
F - TCAACTCCAGGCAGCAGTAAAC & 50.00\% & 54.93 & 28834-\textgreater28855 & \multirow{2}*{118 bps} \\
R - CAAACATTTTGCTCTCAAGCTG & 40.91\% & 51.34 & 28930\textless-28951 & ~  \\ \hline
F - TTCAACTCCAGGCAGCAGTAA & 47.62\% & 52.40 & 28500-\textgreater28520 & \multirow{2}*{128 bps} \\
R - GGCCTTTACCAAACATTTTGC & 42.86\% & 50.45 & 28607\textless-28627 & ~  \\ \hline
F - AATTCAACTCCAGGCAGCAGTAAAC & 44.00\% & 56.04 & 28830-\textgreater28855 & \multirow{2}*{143 bps} \\
R - CCTTGTTGTTGTTGGCCTTTACCAA & 44.00\% & 56.60 & 28948\textless-28973 & ~  \\ \hline
F - CCATTCAGTGCATCGATATCGG & 50.00\% & 53.59 & 28073-\textgreater28097 & \multirow{2}*{200 bps} \\
R - CTGATTTTGGGGTCCATTTAGA & 40.91\% & 50.11 & 28251\textless-28272 & ~  \\ \hline
F - GAGCTATAAAATCAGCACC & 42.11\% & 45.40 & 28012-\textgreater28031 & \multirow{2}*{254 bps} \\
R - TTGGGGTCCATTTAGAGACAT & 42.86\% & 50.31 & 28245\textless-28266 & ~  \\ \hline
\\ \hline
Beta Variant \\ \hline
F - TCATAGCGCTTCCAAAATC & 42.11\% & 47.81 & 25503-\textgreater25522 & \multirow{2}*{911 bps} \\
R - AGACCAGAAGATCAAGAACTCTAG & 41.67\% & 51.29 & 26390\textless-26414 & ~  \\ \hline
F - GTTTGCTAACCCTGTCCTACCAT & 47.83\% & 54.33 & 21740-\textgreater21762 & \multirow{2}*{1,871 bps} \\
R - CTACACCAAGTGACATAGTGTAG & 43.48\% & 50.36 & 23588\textless-23610 & ~  \\ \hline
F - CTACACTATGTCACTTGGTGTA & 40.91\% & 49.16 & 23588-\textgreater23609 & \multirow{2}*{1,927 bps} \\
R - AAGCGCTATGAAAAACAGCAAG & 40.91\% & 52.69 & 25493\textless-25514 & ~  \\ \hline
F - TCATAGCGCTTCCAAAATC & 42.11\% & 47.81 & 25504-\textgreater25522 & \multirow{2}*{3,322 bps} \\
R - CTACTGCTGCCTGGAGTTG & 57.89\% & 52.15 & 28807\textless-28825 & ~  \\ \hline
\\ \hline
Gamma Variant \\ \hline
F - GCCAGAAACCTAAATTGGGTA & 42.86\% & 49.96 & 28102-\textgreater28122 & \multirow{2}*{356 bps} \\
R - CATCTCGACTGCTATTGGTGT & 47.62\% & 52.05 & 28437\textless-28457 & ~  \\ \hline
F - CGAGATGACCAAATTGGCTAC & 47.62\% & 51.34 & 28451-\textgreater28471 & \multirow{2}*{371 bps} \\
R - TTAGAGCTGCCTGGAGTTGAA & 47.62\% & 53.08 & 28801\textless-28821 & ~  \\ \hline
F - ACACCAATAGCAGTCGAGATG & 47.62\% & 52.05 & 28437-\textgreater28457 & \multirow{2}*{385 bps} \\
R - TTAGAGCTGCCTGGAGTTGAA & 47.62\% & 53.08 & 28801\textless-28821 & ~  \\ \hline
\\ \hline
Delta Variant\\ \hline
F - TCAACTCCAGGCAGCAGTATG & 52.38\% & 54.20 & 28807-\textgreater28827  & \multirow{2}*{42 bps} \\
R - CATTCTAGCAGGAGAAGTTCC & 47.62\% & 50.00 & 28828\textless-28848 & ~  \\ \hline
F - GGTAGCAAACCTTGTAATGGT & 42.86\% & 50.30 & 22940-\textgreater22960  & \multirow{2}*{80 bps} \\
R - CCATTAGTGGGTTGGAAACCA & 47.62\% & 52.19 & 22999\textless-23019 & ~  \\ \hline
F - TCTATCAGGCCGGTAGCAAAC & 52.38\% & 54.02 & 22929-\textgreater22949  & \multirow{2}*{101 bps} \\
R - GTAACCAACACCATTAGTGGG & 47.62\% & 50.72 & 23009\textless-23029 & ~  \\ \hline
F - AGGCTTATGAAACTCAAGCCT & 42.86\% & 51.34 & 29342-\textgreater29362  & \multirow{2}*{346 bps} \\
R - AGTGGCCTCGGTGAAAATGTG & 52.38\% & 55.31 & 29667\textless-29687 & ~  \\ \hline
\\ \hline
Omicron Variant\\ \hline
F - CACTCCGCATTACGTTTGGTG & 52.38\% & 54.48 & 27946-\textgreater27966 & \multirow{2}*{56 bps} \\
R - ACCATTCTGGTTACTGCCAGT & 47.62\% & 53.32 & 27981\textless-28001 & ~  \\ \hline
F - ACTCCGCATTACGTTTGGTGG & 52.38\% & 55.42 & 27947-\textgreater27967 & \multirow{2}*{73 bps} \\
R - TTGTTTTGATCGCGCCCCACC & 57.14\% & 58.53 & 27999\textless-28019 & ~  \\ \hline
F - CTCCTTGAAGAATGGAACCT & 45.00\% & 48.79 & 26495-\textgreater26515 & \multirow{2}*{142 bps} \\
R - TTAAAGTTACTGGCCATAACAGCC & 41.67\% & 53.36 & 26613\textless-26637 & ~  \\ \hline
F - GAGCTTAAAAAGCTCCTTGAAG & 40.91\% & 49.90 & 26483-\textgreater26505 & \multirow{2}*{173 bps} \\
R - GCAGCAAGCACAAAACAAGTT & 42.86\% & 53.28 & 26635\textless-26656 & ~  \\ \hline
F - GTGCACAAAAGTTTAACGGCCT & 45.45\% & 52.38 & 24042-\textgreater24063 & \multirow{2}*{312 bps} \\
R - TATGGTTGACCACATCTTGAAG & 40.91\% & 50.23 & 24332\textless-24353 & ~  \\ \hline
\end{tabular}
\caption{Primer pairs successfully validated via in-silico PCR for each SARS-CoV-2 virus variant detection.}
\label{tab:8}
\end{table}
\end{center}
}
}


\bibliographystyle{plain}
\normalem


\begin{appendices}

\section{Data collection and pre-processing} \label{Appendix_1}

In this section, the Table \ref{tab:1} shows the information on the SARS-CoV-2 virus variant labelled by WHO, the Pango lineage, the GISAID clade, the number of samples and the designated label in this project; from the Table \ref{tab:3} can see the average length of each SARS-CoV-2 variant sequence; the Table \ref{tab:2} refer that both SARS-CoV-2 virus and non-SARS-CoV-2 groups collected from GISAID and NCBI, which will be used for calculating the appearance rate of the forward and reverse primers.

\begin{table}[H]
\centering
\begin{tabular}{ccccc}
WHO label & Pango lineage & GISAID clade &  Number of samples & Label\\ \hline
Alpha & B.1.1.7+Q.* & GRY & 119,077 & 1 \\
Beta & B.1.351 & GH/501Y.V2 & 27,782 & 2 \\
Gamma & P.1 & GR/501Y.V3  & 48,588 & 3 \\
Delta & B.1.617.2+AY.* & G/478K.V1  & 142,815 & 4 \\
Omicron & B.1.1.529+BA.* & GR/484A & 135,383 & 0 \\
\hline
\end{tabular}
\caption{SARS-CoV-2 virus variants collected from GISAID database for training the CNN model}
\label{tab:1}

 {\raggedright Note: The SARS-CoV-2 variants are named differently under different classification schemes. The above table lists WHO label, Pango lineage and GISAID clade for five SARS-CoV-2 variants and also the number of samples collected and label used in this project for each variant. \par}

\end{table}

\begin{table}[H]
\centering
\begin{tabular}{ccc}
SARS-CoV-2 virus variant & Average length(base pairs) & Label \\ \hline
Alpha & 29,769 & 1\\
Beta & 29,774 & 2\\
Gamma & 29,770 & 3\\
Delta & 29,766 & 4\\
Omicron & 29,748 & 0\\
\hline
\end{tabular}
\caption{Average gene sequence length for each SARS-CoV-2 virus variant}
\label{tab:3}

 {\raggedright Note: The average gene sequence length of each SARS-CoV-2 variant is quite similar. In this project, the longest sequence is from Delta Variant with 31,079 base-pair. \par}

\end{table}

\begin{table}[H]
\centering
\begin{tabular}{cccc}
Coronavirus species & Source & Host &  Number of samples\\ \hline
SARS-CoV-2(all 5 variants) & GISAID & Homo Sapiens & 58,547 \\
SARS-CoV-2(all 5 variants) & NCBI & Homo Sapiens & 78,692 \\
SARS-CoV-2 & GISAID & Manis javanica & 19 \\
SARS-CoV-2 & GISAID & Rhinolophus affinis & 1\\
SARS-CoV-2 & GISAID & Rhinolophus & 1 \\
SARS-CoV-2 & GISAID & Canis & 29 \\
SARS-CoV-2 & GISAID & Felis catus & 51 \\
MERS-CoV & NCBI & Homo Sapiens & 738 \\
HCoV-OC43 & NCBI & Homo Sapiens & 1,311 \\
HCoV-NL63 & NCBI & Homo Sapiens & 634 \\
HCoV-229E & NCBI & Homo Sapiens & 446 \\
HCoV-HKU1 & NCBI & Homo Sapiens & 404 \\
SARS-CoV-P2 & NCBI & Homo Sapiens & 1 \\
SARS-CoV-HKU-39849 & NCBI & Homo Sapiens & 2 \\
SARS-CoV-GDH-BJH01 & NCBI & Homo Sapiens & 1 \\
HAstV-BF34 & NCBI & Homo Sapiens & 2 \\
\hline
\end{tabular}
\caption{Coronavirus species used for calculating the appearance of the primers}
\label{tab:2}

 {\raggedright Note: In addition to data of SARS-CoV-2 variants, data of Non-human host SARS-CoV-2 virus and other Coronavirus species data are also required when calculating the appearance rate of generated primers. \par}

\end{table}

\section{Feature extracting and evaluating} \label{Appendix_2}

In this section, the Table \ref{tab:10} shows using the different value of top value in top method that obtained primer pairs' number and the range of values for amplicon size; the Table \ref{tab:5} shows the forward primer appearance rates using the Homo Sapiens host; the Table \ref{tab:6} shows the forward primer appearance rates using the non-Homo Sapiens host; the Table \ref{tab:7} shows the percentage of appearance rate of forward primer for other taxa of coronavirus species; the Table \ref{tab:11} and Table \ref{tab:12} show the results of generated reverse primers' number for each forward primer by using 3 different methods for Alpha and Delta variants which are mentioned in Section \ref{Reverse Primer Design} during the reverse primer design; Similar to the table \ref{tab:5}, the Table \ref{tab:5.5} shows the reverse primer appearance rates using the Homo Sapiens host.

\begin{table}[]
\centering
\begin{tabular}{c|cccc}
Variants & Top 75 & Top 125 & Top 175 & Top 250 \\ \hline
Alpha & 60 & 94 & 136 & 171\\
Amplicon Size \textless 1000 & 7 & 11 & 17 & 17\\
Amplicon Size \textless 500 & 7 & 10 & 16 & 15\\
Amplicon Size \textless 300 & 0 & 0 & 0 & 0\\
Amplicon Size \textless 200 & 0 & 0 & 0 & 0\\\hline
 & & & & \\\hline
Delta & 290 & 483 & 623 & 736\\
Amplicon Size \textless 1000 & 0 & 1 & 4 & 2\\
Amplicon Size \textless 500 & 0 & 0 & 0 & 0\\
Amplicon Size \textless 300 & 0 & 0 & 0 & 0\\
Amplicon Size \textless 200 & 0 & 0 & 0 & 0\\
\hline
\end{tabular}
\caption{Total number of primer pairs and the number of primer pairs based on each amplicon size range that was generated according to different top values used in the Alpha and Delta variants }
\label{tab:10}

{\raggedright Note: The results of feature extraction using the top method depend strongly on the set up of top value. \par}

\end{table}

\begin{table}[H]
\centering
\small
\begin{tabular}{c|ccccc}
\multirow{2}*{Forward Primer (5' to 3')} & SARS-CoV-2 & SARS-CoV-2 & SARS-CoV-2 & SARS-CoV-2 & SARS-CoV-2\\
~ &(Alpha) &(Beta) &(Gamma) & (Delta) &(Omicron)\\
\hline
\multirow{2}*{Dataset} & GISAID & GISAID  & GISAID  & GISAID  & GISAID \\
~ & and NCBI & and NCBI & and NCBI & and NCBI & and NCBI\\
Host & Homo Sapiens & Homo Sapiens & Homo Sapiens & Homo Sapiens & Homo Sapiens\\
Sequence Number & 5,000 & 5,000 & 5,000 & 5,000 & 5,000\\ \hline
Alpha Variant & & & & & \\ \hline
TACTAATGATAACACCTCAAG & 0.9928 & 0.0038 & 0.0004 & 0.0 & 0.0 \\
CAATTTGGCAGAGACATTGAT & 0.9928 & 0.0004 & 0.0004 & 0.0 & 0.0 \\
TCAAACTGTCAAACCTGGTAA & 0.9926 & 0.001 & 0.0008 & 0.0002 & 0.0002 \\
CTTTTCAAACTGTCAAACCTG & 0.9924 & 0.001 & 0.0008 & 0.0002 & 0.0002 \\ \hline
Beta Variant & & & & & \\ \hline
CGAACAAACTAAAATGTCTGA & 0.0014 & 0.9832 & 0.0482 & 0.0304 & 0.0016 \\
GCTTAGGGTTGATACAGCCAA & 0.0142 & 0.9756 & 0.0134 & 0.0068 & 0.0056 \\
TAGGGTTGATACAGCCAATCC & 0.0142 & 0.975 & 0.0132 & 0.0068 & 0.0056 \\
AGGGTTGATACAGCCAATCCT & 0.0142 & 0.975 & 0.0132 & 0.0068 & 0.0054 \\ \hline
Gamma Variant & & & & & \\ \hline
TGTGGTAAACAAGCTACACAA & 0.0 & 0.0004 & 0.9958 & 0.0 & 0.0 \\
GTGGTAAACAAGCTACACAAT & 0.0 & 0.0004 & 0.9958 & 0.0 & 0.0 \\
GTGTGGTAAACAAGCTACACA & 0.0 & 0.0004 & 0.9954 & 0.0 & 0.0 \\
ACACAATATCTAGTACAACAG & 0.0002 & 0.0004 & 0.9934 & 0.0 & 0.0 \\ \hline
Delta Variant & & & & & \\ \hline
GATACTAGTTTGTCTGGTTTT & 0.0016 & 0.0382 & 0.0032 & 0.998 & 0.1762 \\
AGTTTGTCTGGTTTTAAGCTA & 0.0016 & 0.0382 & 0.0032 & 0.998 & 0.1766 \\
TATGGTTGATACTAGTTTGTC & 0.0046 & 0.0426 & 0.0082 & 0.9974 & 0.1764 \\
TGGTTGATACTAGTTTGTCTG & 0.0024 & 0.0408 & 0.0056 & 0.9972 & 0.1766 \\ \hline
Omicron Variant & & & & & \\ \hline
GCGCTTCCAAAATCATAACTC & 0.0004 & 0.0002 & 0.0002 & 0.0004 & 0.8344 \\
TCACACCGGAAGCCAATATGG & 0.0002 & 0.0 & 0.0 & 0.0002 & 0.833 \\
AATAACAGTCACACCGGAAGC & 0.0002 & 0.0 & 0.0 & 0.0002 & 0.8328 \\
AGAGATAGGTACGTTAATAGT & 0.0034 & 0.0004 & 0.0006 & 0.0002 & 0.8318 \\
\hline
\end{tabular}
\caption{The frequency of forward primers' appearance (part) tested by 5,000 human (Homo Sapiens) host sequences in different SARS-CoV-2 variants.}
\label{tab:5}

 {\raggedright Note: This is the frequency of appearance of part forward primers of each variant calculated with human host SARS-CoV-2 virus. As seen, these forward primers are specifically generated for one variant only, hence it would hardly appear in other variants. \par}

\end{table}

\begin{sidewaystable}[]
\centering
\begin{tabular}{c|ccccc}
Forward Primer (5' to 3') & SARS-CoV-2 & SARS-CoV-2 & SARS-CoV-2 & SARS-CoV-2 & SARS-CoV-2 \\ \hline
Dataset & GISAID & GISAID & GISAID & GISAID & GISAID\\
Host & Manis javanica & Rhinolophus affinis & Rhinolophus & Canis & Felis catus\\
Sequence Number & 19 & 1 & 1 & 29 & 51 \\ \hline
 \\
 CTCAGACTAATTCTCGTCGGC & 0.0 & 0.0 & 0.0 & 0.0 & 0.0 \\
 \\
 ACTAATTCTCGTCGGCGGGCA & 0.0 & 0.0 & 0.0 & 0.0 & 0.0 \\
 \\
 TCAGACTAATTCTCGTCGGCG & 0.0 & 0.0 & 0.0 & 0.0 & 0.0 \\
 \\
 AGACTAATTCTCGTCGGCGGG & 0.0 & 0.0 & 0.0 & 0.0 & 0.0 \\
 \\
 CAGACTAATTCTCGTCGGCGG & 0.0 & 0.0 & 0.0 & 0.0 & 0.0 \\
 \\
 AACTCCAGGCAGCAGTATGGG & 0.0 & 0.0 & 0.0 & 0.0 & 0.0 \\
 \\
 ACTCCAGGCAGCAGTATGGGA & 0.0 & 0.0 & 0.0 & 0.0 & 0.0 \\
 \\
 CTCCAGGCAGCAGTATGGGAA & 0.0 & 0.0 & 0.0 & 0.0 & 0.0 \\
 \\
 CCAGGCAGCAGTATGGGAACT & 0.0 & 0.0 & 0.0 & 0.0 & 0.0 \\
 \\
 AGGCAGCAGTATGGGAACTTC & 0.0 & 0.0 & 0.0 & 0.0 & 0.0 \\
 \\
\hline
\end{tabular}
\caption{The frequency of forward primers' appearance (part) tested by non-human (non-Homo Sapiens) host sequences in SARS-CoV-2 virus.}
\label{tab:6}

 {\raggedright Note: This is the frequency of appearance of part forward primers for Delta variant calculated with non-human host SARS-CoV-2 virus. As seen, these forward primers are specifically generated for the Delta variant with human host, hence the appearance rates are 0.0\%. \par}

\end{sidewaystable}

\begin{sidewaystable}[]
\centering
\begin{tabular}{c|ccccccccc}
\multirow{2}*{Forward Primer (5' to 3')} & \multirow{2}*{MERS-CoV} & \multirow{2}*{HCoV-OC43} & \multirow{2}*{HCoV-NL63} & HCoV & HCoV & SARS-CoV & SARS-CoV & SARS-CoV & HAstV\\
~ & ~ & ~ & ~ & -229E & -HKU1 & -P2 & -HKU-39849 & -GDH-BJH01 & -BF34 \\ \hline
Dataset & NCBI & NCBI & NCBI & NCBI & NCBI & NCBI & NCBI & NCBI & NCBI\\
Host: HS (Homo Sapiens) & HS & HS & HS & HS & HS & HS  & HS & HS & HS \\
Sequence Number & 738 & 1,311 & 634 & 446 & 404 & 1 & 2 & 1 & 2 \\ \hline
\\
CTCAGACTAATTCTCGTCGGC & 0.0 & 0.0 & 0.0 & 0.0 & 0.0 & 0.0 & 0.0 & 0.0 & 0.0\\
\\
ACTAATTCTCGTCGGCGGGCA & 0.0 & 0.0 & 0.0 & 0.0 & 0.0 & 0.0 & 0.0 & 0.0 & 0.0\\
\\
TCAGACTAATTCTCGTCGGCG & 0.0 & 0.0 & 0.0 & 0.0 & 0.0 & 0.0 & 0.0 & 0.0 & 0.0\\
\\
AGACTAATTCTCGTCGGCGGG & 0.0 & 0.0 & 0.0 & 0.0 & 0.0 & 0.0 & 0.0 & 0.0 & 0.0\\
\\
CAGACTAATTCTCGTCGGCGG & 0.0 & 0.0 & 0.0 & 0.0 & 0.0 & 0.0 & 0.0 & 0.0 & 0.0\\
\\
AACTCCAGGCAGCAGTATGGG & 0.0 & 0.0 & 0.0 & 0.0 & 0.0 & 0.0 & 0.0 & 0.0 & 0.0\\
\\
ACTCCAGGCAGCAGTATGGGA & 0.0 & 0.0 & 0.0 & 0.0 & 0.0 & 0.0 & 0.0 & 0.0 & 0.0\\
\\
CTCCAGGCAGCAGTATGGGAA & 0.0 & 0.0 & 0.0 & 0.0 & 0.0 & 0.0 & 0.0 & 0.0 & 0.0\\
\\
CCAGGCAGCAGTATGGGAACT & 0.0 & 0.0 & 0.0 & 0.0 & 0.0 & 0.0 & 0.0 & 0.0 & 0.0\\
\\
AGGCAGCAGTATGGGAACTTC & 0.0 & 0.0 & 0.0 & 0.0 & 0.0 & 0.0 & 0.0 & 0.0 & 0.0\\
\\
\hline
\end{tabular}
\caption{The frequency of forward primers' appearance (part) tested by human (Homo Sapiens) host sequences in different Coronavirus species.}
\label{tab:7}

 {\raggedright Note: This is the frequency of appearance of part forward primers for Delta variant calculated with Coronavirus species. As seen, these forward primers are specifically generated for the Delta variant only, hence the appearance rates are 0.0\%. \par}

\end{sidewaystable}

\begin{table}[H]
\centering
\small
\begin{tabular}{c|c|c|c}
Forward Primer (5' to 3') & Alpha & Alpha & Alpha \\
 & Pooling Method & Top method & Mix Method \\ \hline
TTGGCAGAGACATTGATGACA & 30 & 96 & 58\\ \hline
TCTTATGGGTTGGGATTATCC & 48 & 61	& 115\\ \hline
TTGCACGTCTTGACAAAGTTG & 37 & 60	& 52\\ \hline
TCCTTGCACGTCTTGACAAAG & 40 & 51	& 43\\ \hline
TGCACGTCTTGACAAAGTTGA & 34 & 63	& 38\\ \hline
TTTGGCAGAGACATTGATGAC & 47 & 84	& 56\\ \hline
CACAACACATTTGTGTCTGGT & 26 & 53	& 38\\ \hline
CTTGCACGTCTTGACAAAGTT & 42 & 57	& 49\\ \hline
CACACAACACATTTGTGTCTG & 21 & 54	& 36\\ \hline
ACACAACACATTTGTGTCTGG & 32 & 61	& 44\\ \hline
CCTTGCACGTCTTGACAAAGT & 33 & 58	& 38\\ \hline
GCACGTCTTGACAAAGTTGAG & 32 & 22	& 45\\ \hline
Average Number & 35.166 & 52.000 & 51.000 \\
\hline
\end{tabular}
\caption{Reverse primers generated by 3 different methods for each Alpha variant-specific forward primer.}
\label{tab:11}

 {\raggedright Note: Depending on the method of feature extraction chosen, the final number of primes obtained will differ. This table shows the number of reverse primer generated by using 3 different extraction methods for Alpha variant. \par}

\end{table}

\begin{table}[H]
\centering
\small
\begin{tabular}{c|c|c|c}
Forward Primer (5' to 3') &	Delta &	Delta &	Delta \\
 & Pooling Method & Top method & Mix Method \\ \hline
GATCACCGGTGGAATTGCTAC	& 16	& 68	& 29 \\ \hline
GAATTGCTACCGCAATGGCTT	& 15	& 55	& 25 \\ \hline
ACTCAGACTAATTCTCGTCGG	& 29	& 62	& 35 \\ \hline
CTACCGCAATGGCTTGTCTTG	& 18	& 49	& 23 \\ \hline
TGGAATTGCTACCGCAATGGC	& 17	& 60	& 29 \\ \hline
CTCAGACTAATTCTCGTCGGC	& 27	& 70	& 41 \\ \hline
ATTGCTACCGCAATGGCTTGT	& 20	& 60	& 24 \\ \hline
AGACTCAGACTAATTCTCGTC	& 27	& 60	& 42 \\ \hline
TAGATTTTGTTCGCGCTACTG	& 18	& 53	& 30 \\ \hline
CCGGTGGAATTGCTACCGCAA	& 16	& 70	& 23 \\ \hline
ACCGCAATGGCTTGTCTTGTA	& 17	& 74	& 22 \\ \hline
GGTGGAATTGCTACCGCAATG	& 16	& 63	& 28 \\ \hline
CTCCTTTAGATTTTGTTCGCG	& 26	& 65	& 16 \\ \hline
TCACCGGTGGAATTGCTACCG	& 7	    & 60	& 24 \\ \hline
ACCGGTGGAATTGCTACCGCA	& 18	& 54	& 24 \\ \hline
ATCACCGGTGGAATTGCTACC	& 20	& 59	& 25 \\ \hline
TACCGCAATGGCTTGTCTTGT	& 22	& 67	& 18 \\ \hline
GGAATTGCTACCGCAATGGCT	& 14	& 66	& 28 \\ \hline
CCGCAATGGCTTGTCTTGTAG	& 21	& 61	& 31 \\ \hline
GTGGAATTGCTACCGCAATGG	& 15	& 64	& 22 \\ \hline
TTGCTACCGCAATGGCTTGTC	& 16	& 66	& 31 \\ \hline
GCTACCGCAATGGCTTGTCTT	& 10	& 66	& 27 \\ \hline
TCAGACTCAGACTAATTCTCG	& 27	& 67	& 47 \\ \hline
AATTGCTACCGCAATGGCTTG	& 15	& 60	& 29 \\ \hline
CAGACTCAGACTAATTCTCGT	& 18	& 77	& 39 \\ \hline
CGGTGGAATTGCTACCGCAAT	& 14	& 57	& 24 \\ \hline
TGCTACCGCAATGGCTTGTCT	& 19	& 72	& 16 \\ \hline
GACTCAGACTAATTCTCGTCG	& 26	& 74	& 44 \\ \hline
Average Number & 18.714 & 63.536 & 28.429 \\
\hline
\end{tabular}
\caption{The number of reverse primer generated by 3 different methods for each Delta variant forward primer.}
\label{tab:12}

 {\raggedright Note: Depending on the method of feature extraction chosen, the final number of primes obtained will differ. This table shows the number of reverse primer generated by using 3 different extraction methods for Delta variant. \par}

\end{table}

\begin{table}[H]
\centering
\small
\begin{tabular}{c|ccccc}
\multirow{2}*{Reverse Primer (5' to 3')} & SARS-CoV-2 & SARS-CoV-2 & SARS-CoV-2 & SARS-CoV-2 & SARS-CoV-2\\
~ &(Alpha) &(Beta) &(Gamma) & (Delta) &(Omicron)\\
\hline
\multirow{2}*{Dataset} & GISAID & GISAID  & GISAID  & GISAID  & GISAID \\
~ & and NCBI & and NCBI & and NCBI & and NCBI & and NCBI\\
Host & Homo Sapiens & Homo Sapiens & Homo Sapiens & Homo Sapiens & Homo Sapiens\\
Sequence Number & 5,000 & 5,000 & 5,000 & 5,000 & 5,000\\ \hline
Alpha Variant & & & & & \\ \hline
CATCAATGTCTCTGCCAAATTG & 0.9936 & 0.0004 & 0.0008 & 0.0002 & 0.0 \\
ACCAGACACAAATGTGTTGTGT & 0.9928 & 0.0008 & 0.0018 & 0.0004 & 0.0 \\
CAGACACAAATGTGTTGTGTG & 0.992 & 0.0008 & 0.0018 & 0.0004 & 0.0 \\
ACAGCATCAGTAGTGTCATCA & 0.9918 & 0.0004 & 0.0008 & 0.0002 & 0.0 \\ \hline
Beta Variant & & & & & \\ \hline
ACAGGGTTAGCAAACCTCTT & 0.0 & 0.9638 & 0.0 & 0.0006 & 0.0 \\
CTACTGCTGCCTGGAGTTG & 0.0016 & 0.9594 & 0.0004 & 0.0006 & 0.0008 \\
GTTCGTTTAGACCAGAAGATCAAG & 0.0002 & 0.9546 & 0.0 & 0.0004 & 0.0 \\
GGTTATGATTTTGGAAGCGCTA & 0.0006 & 0.953 & 0.0178 & 0.0012 & 0.0004 \\ \hline
Gamma Variant & & & & & \\ \hline
AATTTGGTCATCTCGACTG & 0.0 & 0.0 & 0.9918 & 0.0002 & 0.0 \\
TGGTCATCTCGACTGCTATTGGTGT & 0.0 & 0.0 & 0.9914 & 0.0002 & 0.0 \\
GCCAATTTGGTCATCTCGAC & 0.0 & 0.0 & 0.9912 & 0.0002 & 0.0 \\
TGAACCGTCGATTGTGTGAA & 0.0& 0.0006 & 0.985 & 0.0002 & 0.0 \\ \hline
Delta Variant & & & & & \\ \hline
CATTGCGGTAGCAATTCCA & 0.0006 & 0.0002 & 0.0 & 0.9952 & 0.165 \\
AGCGCGAACAAAATCTAAAGGA & 0.001 & 0.003 & 0.0006 & 0.9934 & 0.1634 \\
GTAGCGCGAACAAAATCTAAAGGAG & 0.0008 & 0.003 & 0.0006 & 0.9932 & 0.163 \\
GACGAGAATTAGTCTGAGTCTGAT & 0.0032 & 0.0004 & 0.0004 & 0.9896 & 0.1624 \\ \hline
Omicron Variant & & & & & \\ \hline
ATTGTGCCAACCACCATAGAA & 0.0038 & 0.0008 & 0.002 & 0.0036 & 0.834 \\
GGTGTGACTGTTATTGCCTGACCA & 0.0 & 0.0 & 0.0 & 0.0 & 0.8292 \\
TAACGTACCTATCTCTTCCGAA & 0.0032 & 0.0 & 0.001 & 0.0 & 0.8286 \\
CACCTGTGCCTTTTAAACCATTG & 0.0 & 0.0 & 0.0 & 0.0002 & 0.828 \\
\hline
\end{tabular}
\caption{The frequency of SARS-CoV-2 variant-specific reverse primers appearance evaluated on 5,000 human (Homo Sapiens)}
\label{tab:5.5}

 {\raggedright Note: This is the frequency of appearance of part reverse primers of each variant calculated with human host SARS-CoV-2 virus. As seen, these reverse primers are also specifically generated for one variant only, hence it would hardly appear in other variants. \par}

\end{table}

\section{Flowchart of the methodology}

\begin{figure}[H]
\centering
\includegraphics[width=0.8 \textwidth]{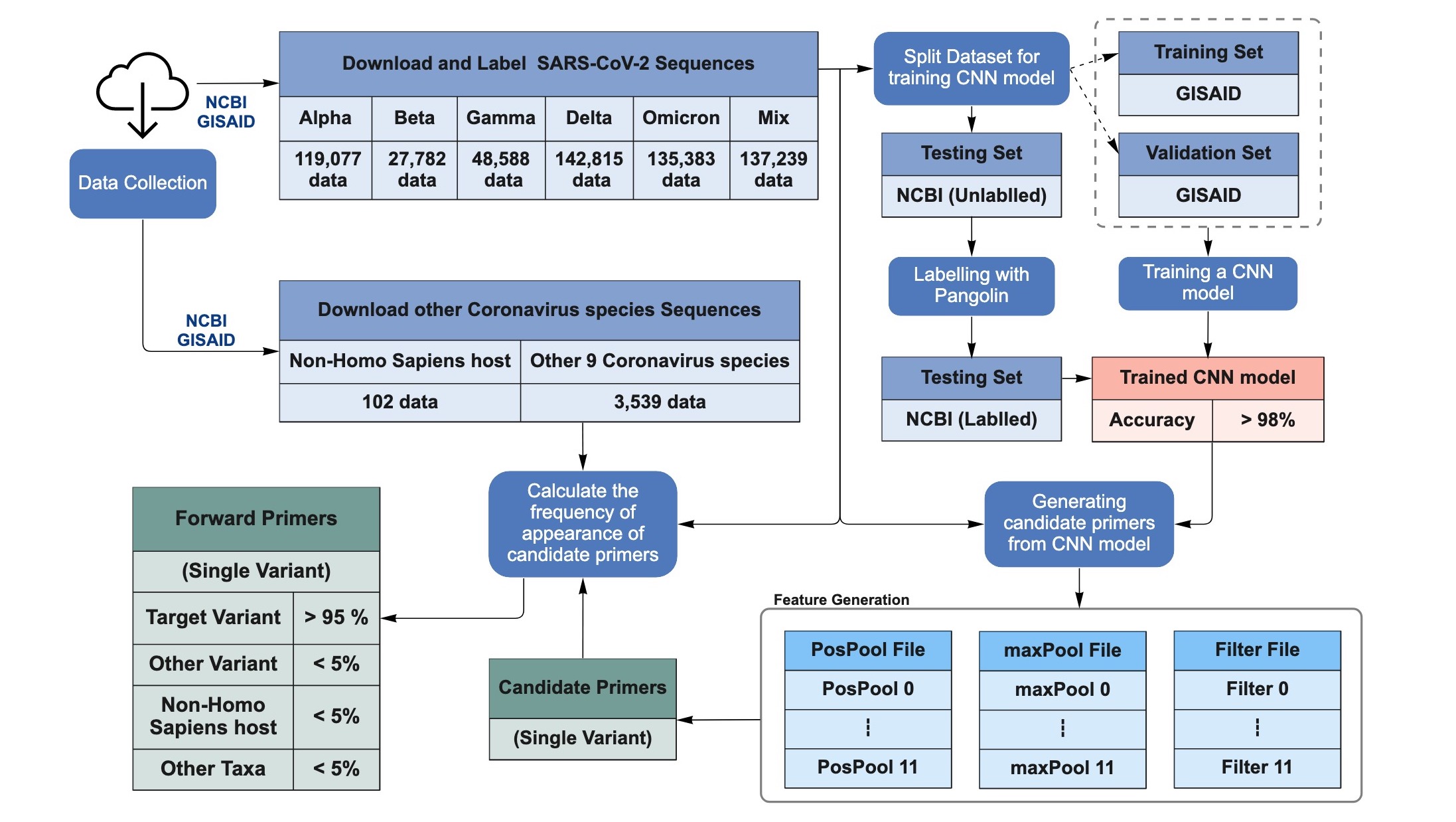}
    \caption{\label{Flowchart_Forward}The flowchart of satisfactory forward primer design.}
\end{figure}

\begin{figure}[H]
\centering
\includegraphics[width=0.8 \textwidth]{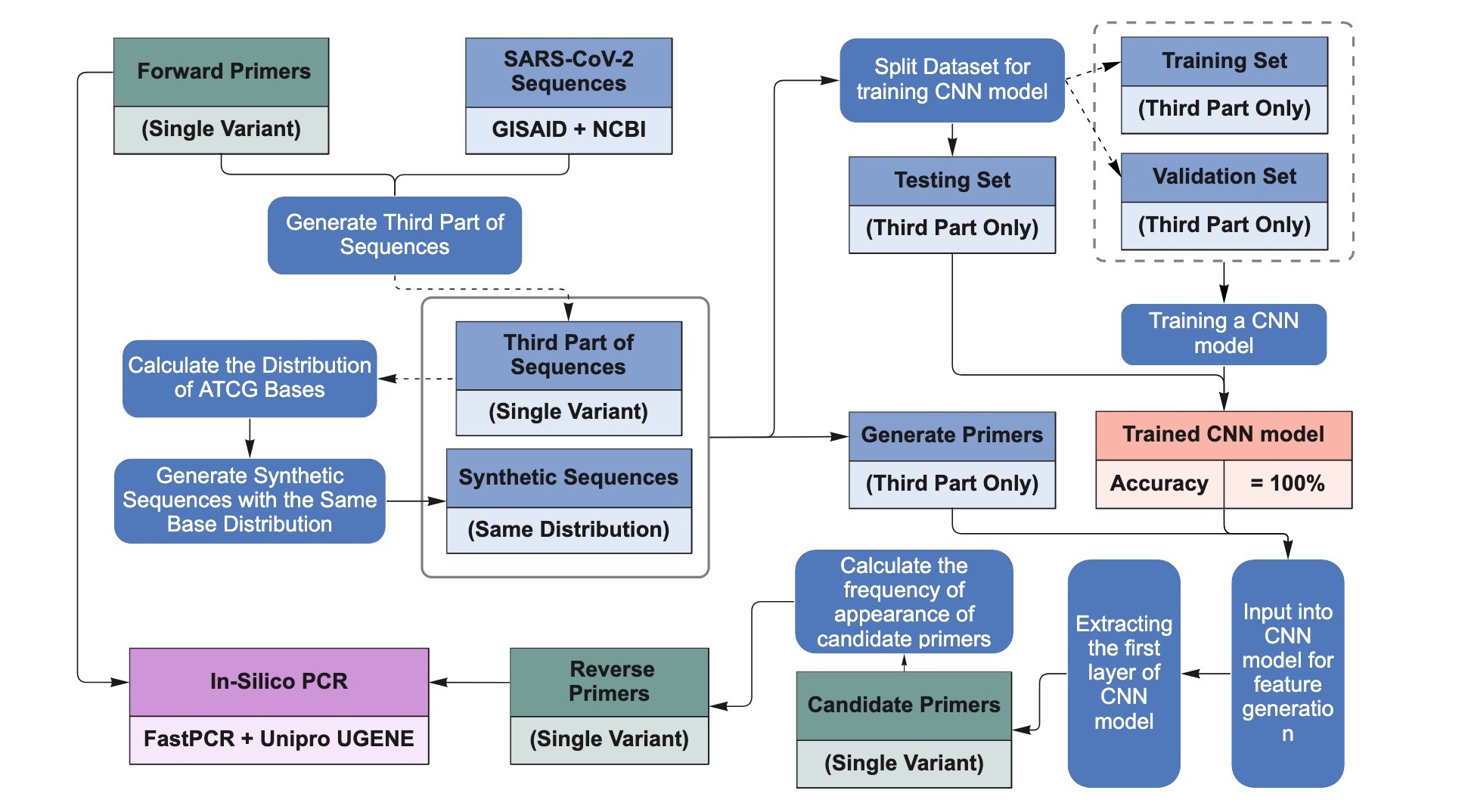}
    \caption{\label{Flowchart_Reverse}The flowchart of the reverse primer design.}
\end{figure}

\begin{figure}[H]
\centering
\includegraphics[width=0.8 \textwidth]{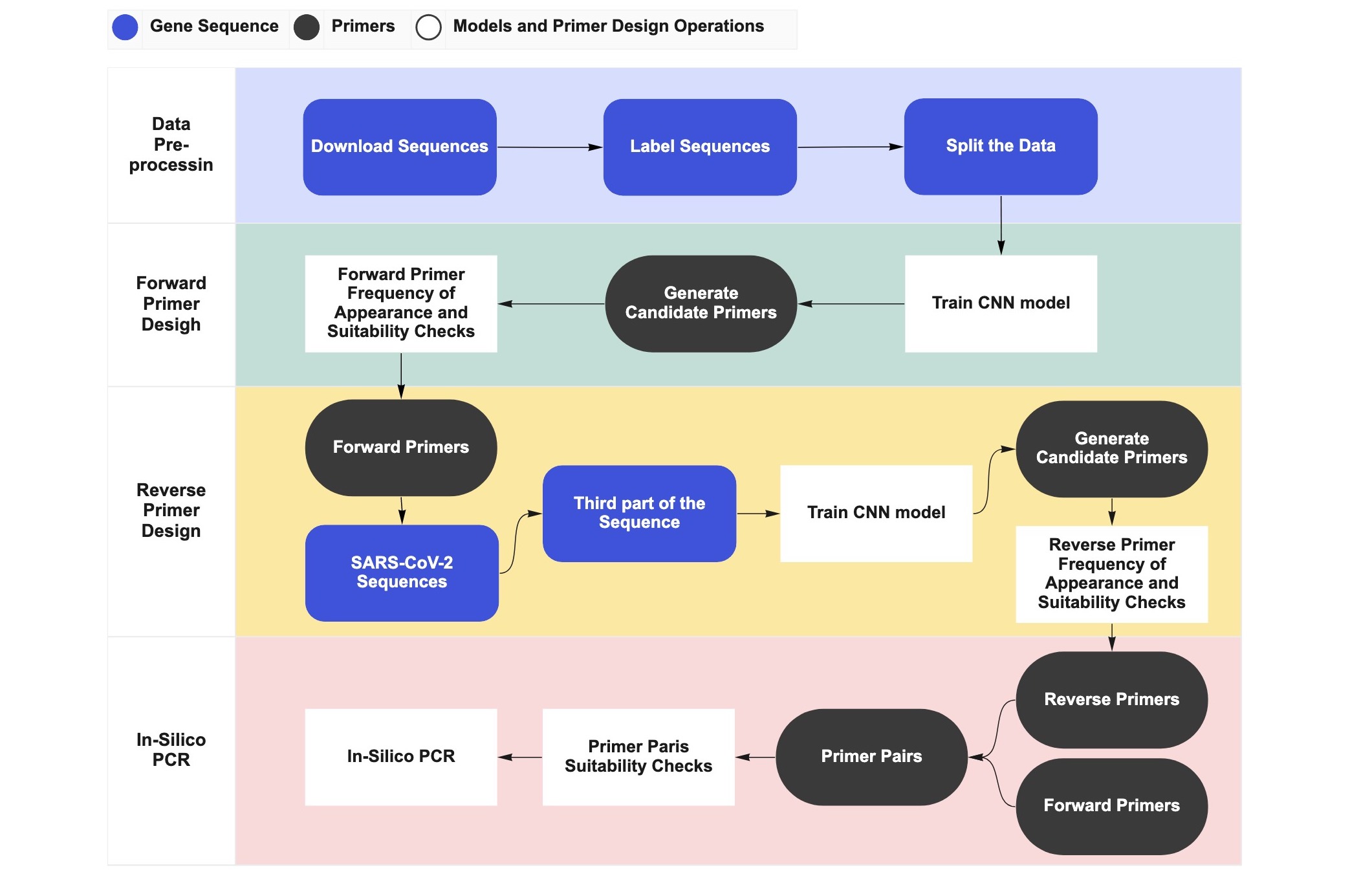}
    \caption{\label{Flowchart_All}The flowchart to show the overall forward and reverse primer design workflow.}
\end{figure}

In this section, the Fig \ref{Flowchart_Forward} and Fig \ref{Flowchart_Reverse} show the whole process of forward primer design and reverse primer design. The Fig \ref{Flowchart_Forward} includes the entire process of forward primer design from data collection to feature extraction and the final candidate forward primer generation after evaluating; while the following Fig \ref{Flowchart_Reverse} which shows the reverse primer design completed afterwards forward primer design; the Fig \ref{Flowchart_All} gives an outline of the whole process of the approach.

\section{BLAST and in-silico PCR for primer validation} \label{Appendix_4}

\begin{figure}[H]
\centering
\includegraphics[width=0.4 \textwidth]{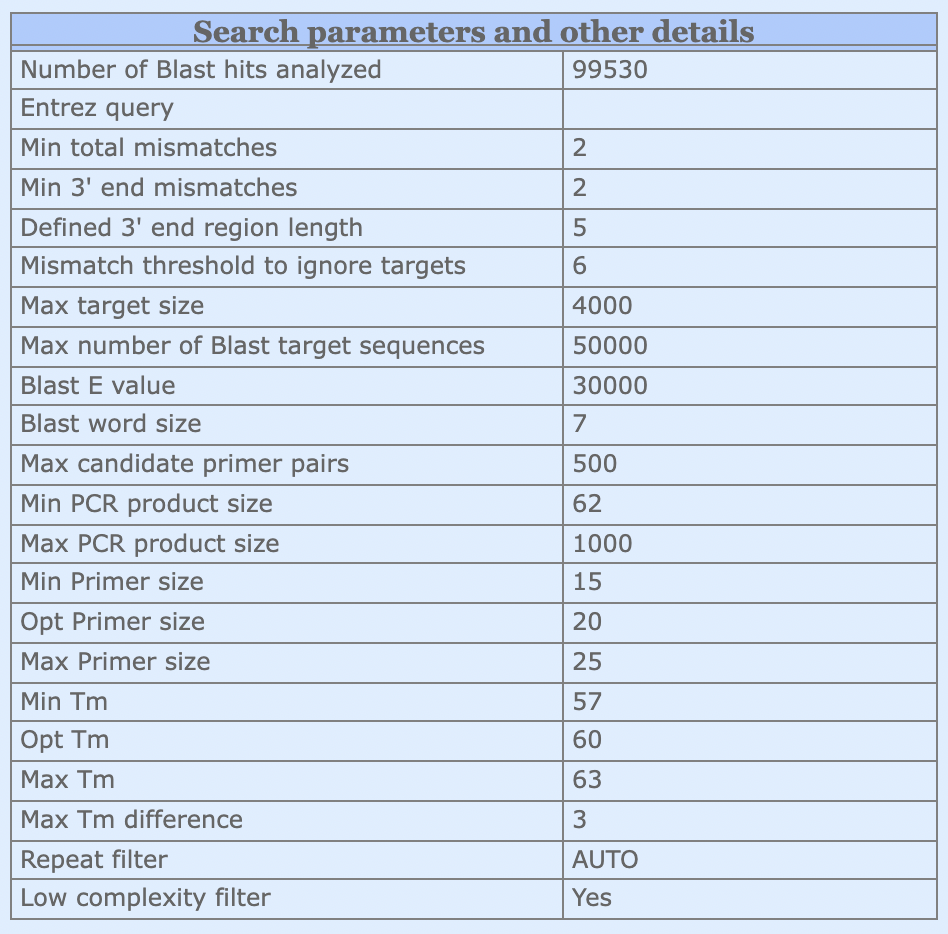}
\caption{BLAST search parameters and other details.}
\label{fig:BALST}

 {\raggedright Note: Performing Primer-BLAST analysis of the primer pairs obtained by deep-CNN model on the NCBI website. There are some parameters that need to be set before performing BLAST to ensure the validity of the analysis. \par}

\end{figure}

\begin{figure}[H]
\centering
\includegraphics[width=0.8 \textwidth]{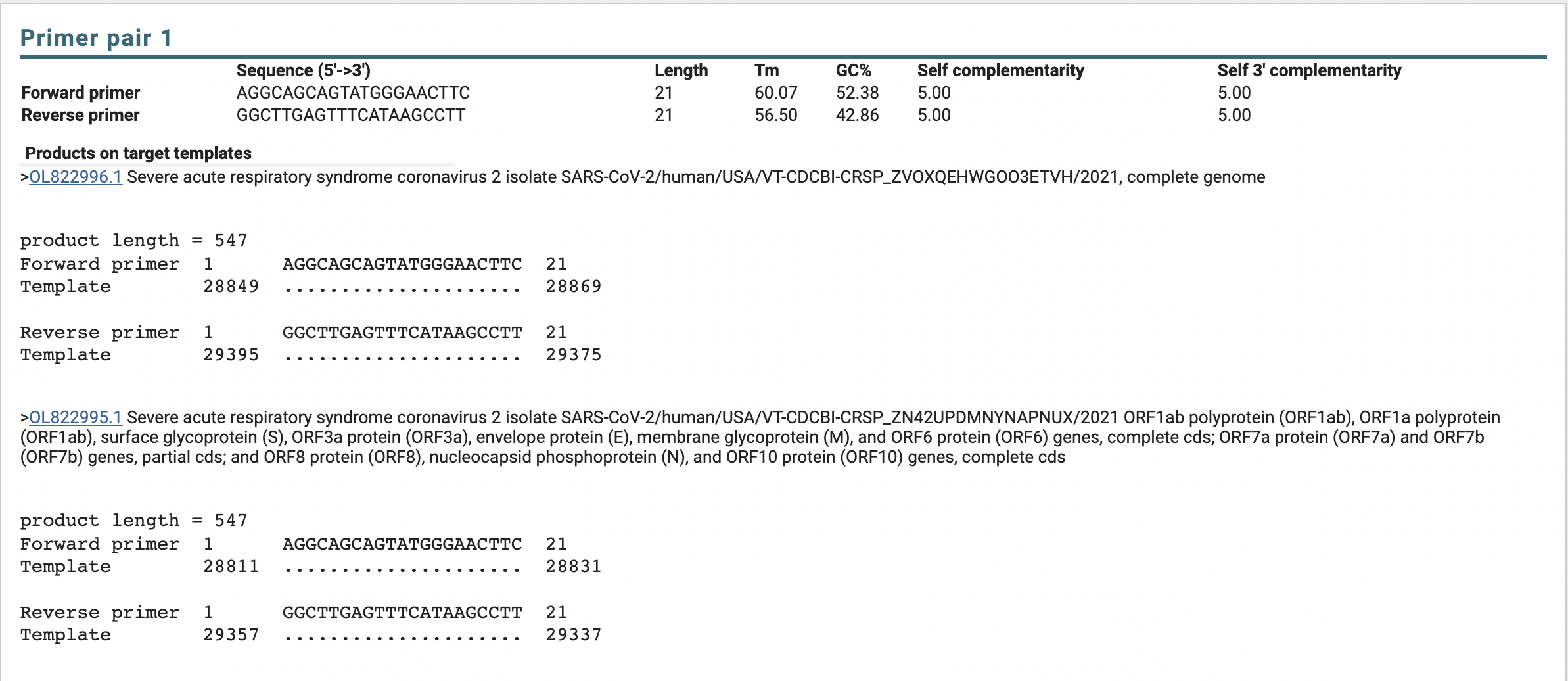}
\caption{Snapshot of primer specificity checks using Primer-BLAST.}
\label{fig:BLAST_result}

 {\raggedright Note: This is the result of a Primer-BLAST completed on a primer pair, showing the Accession\_ID and species of the gene sequence for which the primer pair could be searched, and the amplicon size of the PCR performed in that gene sequence using this primer pair. \par}

\end{figure}

In this section, the Fig \ref{fig:BALST} and Fig \ref{fig:BLAST_result} show the result of Primer-BLAST analysis to confirm that the selected primer pair would specifically target SARS-CoV-2 virus.

\begin{figure}[H]
\centering
\includegraphics[width=0.63 \textwidth]{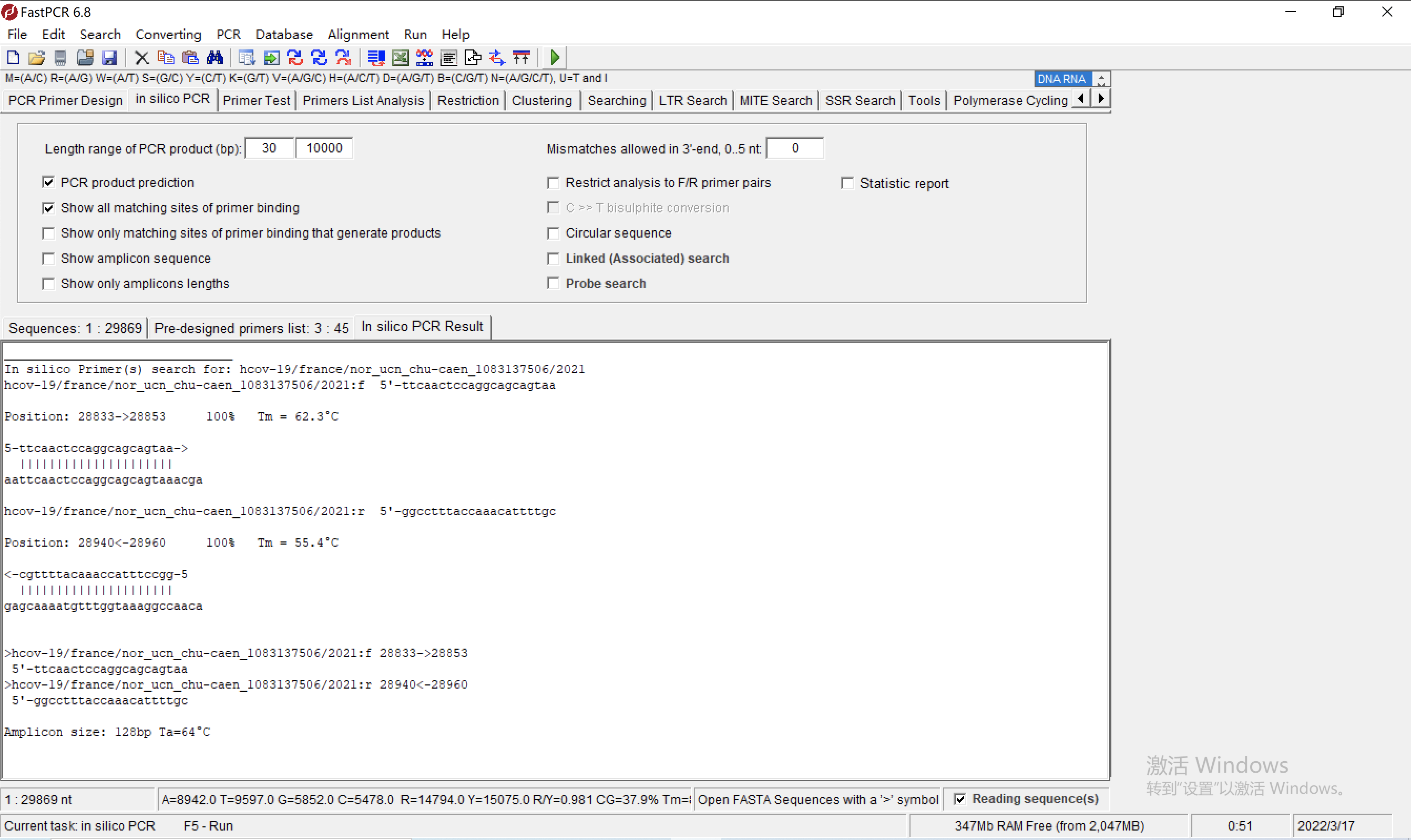}
\caption{Snapshot of in-silico PCR analysis of chosen primer pairs using FastPCR software.}
\label{fig:FastPCR}
\end{figure}

\begin{figure}[H]
\centering
\includegraphics[width=0.7 \textwidth]{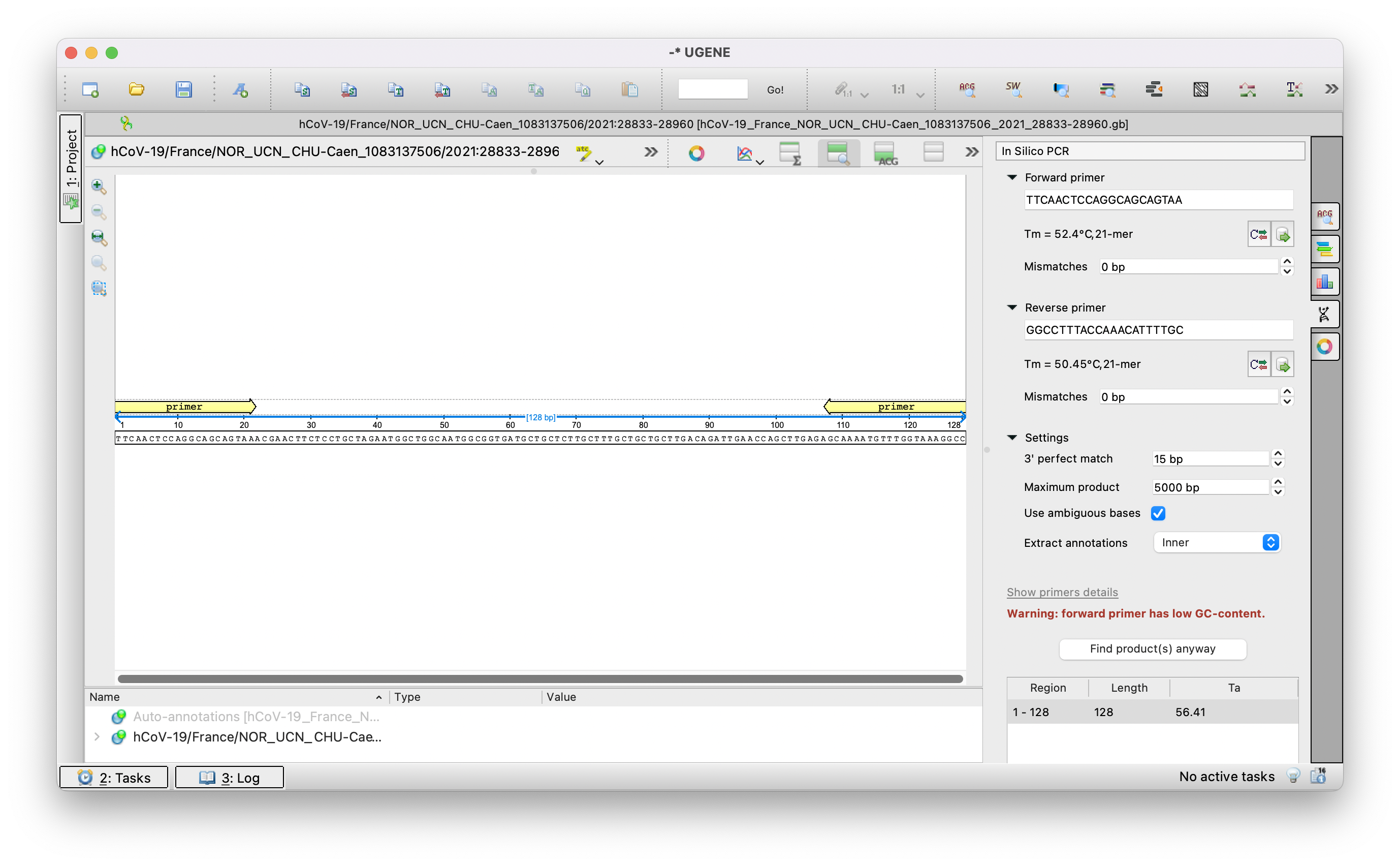}
\caption{Snapshot of in-silico PCR analysis of pre-selected primer pairs using the Unipro UGENE software.}
\label{fig:Unipro_UGENE}
\end{figure}

In this section, the Fig \ref{fig:FastPCR} and Fig \ref{fig:Unipro_UGENE} show the result of in-silico PCR by using FastPCR software and Unipro UGENE software.

\section{Compare with Primer3Plus} \label{Appendix_3}

In this section, the Table \ref{tab:14} shows the result of primer pairs generated by using our deep-learning method for both the forward and reverse primers and compare it with the result using Primer3Plus.

\begin{table}[H]
\centering
\small
\begin{tabular}{ccccc}
\multicolumn{1}{c}{} & \multicolumn{2}{c}{\textbf{Deep-learning Method}} & \multicolumn{2}{c}{\textbf{Primer3Plus}} \\ \hline
\multicolumn{1}{c}{} & \multicolumn{2}{c}{(Flexible with no upper limit)} & \multicolumn{2}{c}{(Up to 10000 base pairs)} \\
\textbf{Forward Primer} & \multicolumn{4}{c}{} \\ \hline
\multicolumn{5}{c}{} \\
\multicolumn{1}{c|}{\textit{Total at the beginning}} & 3,626 primers  & \multicolumn{1}{l|}{} & 30 primers & \\
\multicolumn{1}{c|}{\textit{GC content check}} & 2,725 primers & \multicolumn{1}{l|}{$\downarrow$24.85\%} & 30 primers & \multicolumn{1}{l}{$\downarrow$0\%} \\
\multicolumn{1}{c|}{\textit{Frequency of appearance check}} & 29 primers & \multicolumn{1}{l|}{$\downarrow$98.94\%}  & 0 primers & \multicolumn{1}{l}{$\downarrow$100\%} \\
\multicolumn{5}{c}{} \\ \hline
\textbf{Reverse Primer} & \multicolumn{4}{c}{} \\ \hline
\multicolumn{5}{c}{} \\
\multicolumn{1}{c|}{\textit{Total at the beginning}} & 53,777 primers  & \multicolumn{1}{c|}{} & 30 primers & \\
\multicolumn{1}{c|}{\textit{GC content check}} & 34,955 primers & \multicolumn{1}{l|}{$\downarrow$35\%} & 30 primers & \multicolumn{1}{l}{$\downarrow$0\%} \\
\multicolumn{1}{c|}{\textit{Frequency of appearance check}} & 357 primers & \multicolumn{1}{l|}{$\downarrow$98.98\%}  & 0 primers & \multicolumn{1}{l}{$\downarrow$100\%} \\
\multicolumn{5}{c}{} \\ \hline
\textbf{Primer Pairs} & \multicolumn{4}{c}{} \\ \hline
\multicolumn{5}{c}{} \\
\multicolumn{1}{c|}{\textit{Primer Design Rules}} & \multicolumn{2}{c|}{143 primer paris} & \multicolumn{2}{c}{0 primer pairs} \\
\multicolumn{5}{c}{} \\ \hline
\end{tabular}
\caption{The comparison with the primer results of deep-learning method and Primer3Plus}
\label{tab:14}

 {\raggedright Note: Since Primer3Plus can only input sequences of up to 10,000 base pairs at one time and generates 10 primer pairs (10 forward + 10 reverse) for every input. For a SARS-CoV-2 virus with an average length of 30,000 base pairs, the sequence needs to be divided into 3 segments and input separately, resulting in a total of 30 primer pairs. \par}

\end{table}

\end{appendices}

\end{document}